\documentclass[lettersize,journal]{IEEEtran}
\usepackage{amsmath,amsfonts}

\usepackage{algorithm}
\usepackage{array}
\usepackage[caption=false,font=normalsize,labelfont=sf,textfont=sf]{subfig}
\usepackage{textcomp}
\usepackage{stfloats}
\usepackage{url}
\usepackage{verbatim}
\usepackage{graphicx}
\usepackage{tikz}
\usepackage{cite}
\usepackage{titlesec}
\usepackage{xcolor}
\usepackage{pifont}
\usepackage{flushend}
\usepackage[latin1]{inputenc}
\usepackage{colortbl}
\usepackage{soul}
\usepackage{multirow}
\usepackage{pifont}
\usepackage{color}
\usepackage{alltt}
\usepackage[hidelinks]{hyperref}
\usepackage{enumerate}
\usepackage{siunitx}
\usepackage{breakurl}
\usepackage{epstopdf}
\usepackage{pbox}
\usepackage{array} 
\usepackage{algpseudocode}
\usepackage{booktabs} 
\usepackage{booktabs, multirow, array}
\usepackage{amsmath}  
\usepackage{amsthm}   

\newtheorem{theorem}{Theorem}

\newcommand{\hx}[2]{\textcolor{blue}{#1}}

\titlespacing{\section}{0pt}{*0.7}{*0.7}
\titlespacing{\subsection}{0pt}{*0.5}{*0.5}

\begin{document}

\title{DT-MPC: Synthesizing Derivation-Free Model Predictive Control from Power Converter Netlists via Physics-Informed Neural Digital Twins}


\author{
	\vskip 1em
	Jialin Zheng, \emph{Member,~IEEE},
	Haoyu Wang, \emph {Graduate Student Member,~IEEE},
    Yangbin Zeng, \emph{Member,~IEEE},\\
    Han Xu, \emph{Student Member,~IEEE},
    Di Mou, \emph{Member,~IEEE},
    Hong Li, \emph{Senior Member,~IEEE},\\
    Patrick Wheeler, \emph{Fellow,~IEEE},
    Sergio Vazquez, \emph{Fellow,~IEEE},
    Leopoldo G. Franquelo, \emph{Life Fellow,~IEEE}
        \vspace{-15pt}

	\thanks{
	
		  Jialin Zheng and Haoyu Wang contributed equally to this work. 
       
		This paper was supported by the National Natural Science Foundation of China under Grants 52577197 and 52237008, the Guangdong Basic and Applied Basic Research Foundation under Grant 2025A1515011715, and Project PID2023-1522920B-100 funded by MCIN/AEI 10.13039/501100011033 and by ERDF/EU. \emph{(Corresponding author: Yangbin Zeng)}.
	}
}

\markboth{Journal of \LaTeX\ Class Files,~Vol.~14, No.~8, August~2025}%
{Shell \MakeLowercase{\textit{et al.}}: A Sample Article Using IEEEtran.cls for IEEE Journals}


\maketitle

\begin{abstract}
Model Predictive Control (MPC) is a powerful control strategy for power electronics, but it highly relies on manually-derived and topology-specific analytical models, which is labor-intensive and time-consuming in practical designs. To overcome this bottleneck, this paper introduces a Digital-Twin-based MPC (DT-MPC) framework for generic power converters that can systematically translate a high-level circuit into an objective-aware control policy by leveraging a DT as a high-fidelity system model. Furthermore, a physics-informed neural surrogate predictor is proposed to accelerate predictions by DT and enable real-time operation. A gradient-free simplex search optimizer is also introduced to efficiently handle complex multi-objective optimization. The efficacy of the framework has been validated through a cloud-to-edge deployment on a 1500 W dual active bridge converter. Experimental results show that the synthesized predictive model achieves an inference speed over 7 times faster than real time, the DT-MPC controller outperforms several human-designed counterparts, and the overall framework reduces engineering design time by over 95\%, verifying the superiority of DT-MPC on generalized power converters.
\end{abstract} 

\begin{IEEEkeywords}
Digital twins, model predictive control, DC-DC converter, neural network, simplex optimization.
\end{IEEEkeywords}

\markboth{IEEE TRANSACTIONS ON Power ELECTRONICS}%
{}

\definecolor{limegreen}{rgb}{0.2, 0.8, 0.2}
\definecolor{forestgreen}{rgb}{0.13, 0.55, 0.13}
\definecolor{greenhtml}{rgb}{0.0, 0.5, 0.0}

\section{Introduction}

\IEEEPARstart{M}{odel} Predictive Control (MPC) has gained significant attention as an advanced control technique for power electronic converters and motor drives. It reframes a control problem as an explicit forward-looking online optimization task \cite{6775336}, enabling it to intuitively handle multi-variable coupling, physical constraints, and strong nonlinearities where traditional linear control methods often fall short \cite{6317184}. With the exponential growth of processing power in embedded computing units, MPC is transitioning from a theoretical concept to widespread applications \cite{9736455,8972584, 10159491,9209111}. However, the deployment of its powerful capabilities is constrained by the critical reliance on manually-derived, topology-specific analytical models. Such models are not only difficult to generalize across different converter architectures but also impose a significant engineering burden on controller design and implementation.

These challenges are particularly evident in high-frequency isolated DC-DC converter applications, such as Dual-Active Bridge (DAB) and Multi-Active Bridge (MAB) converters \cite{zvs_mab, cps_mab, ups_mab, time_mab, reactive_mab, ecce_mab}, where achieving effective control requires precise characterization of rapidly switching, mode-dependent dynamics. These converters are characterized by complex systems and wide-ranging dynamics, and their microsecond-level switching periods impose extreme real-time constraints on control algorithms \cite{9422148}. Concurrently, the continuous nature of their modulation schemes not only requires the analytical model to handle inputs over a continuous range but also presents the online optimizer with an infinite set of candidate actions, making the optimization problem exceptionally complex \cite{8960628, 10168292}. These challenges collectively make the design of analytical-model-based MPC exceedingly difficult in high-frequency converters.

To overcome these challenges and pursue higher performance within the analytical model framework, several strategies have been proposed to enhance both the transient and steady-state performance. Numerous MPC schemes have been developed specifically for DAB and MAB converters, focusing on modulation techniques such as Single-Phase Shift (SPS) \cite{8717683, 7448412}, Dual-Phase Shift (DPS) \cite{8799001, 9940589}, and Triple-Phase Shift (TPS) \cite{8945348, 10315245, 8799001}. For instance, to achieve a fast transient response, a dead-beat control strategy was proposed, but this method relies on high-bandwidth current sensors \cite{7448412}. To improve the efficiency of DAB converters, an MPC strategy based on DPS modulation was introduced; however, it struggles to simultaneously achieve minimum current stress and full-range Zero-Voltage Switching (ZVS), requiring a complex cost function to balance these conflicting objectives \cite{9940589}. To overcome this inherent trade-off, research has progressed to more advanced TPS strategies. The additional degrees of freedom (DOFs) in TPS enable the simultaneous optimization of ZVS performance alongside other objectives like power flow and current stress, as demonstrated in \cite{10315245}. However, this enhanced capability introduces greater modeling complexity. Subsequent studies have therefore focused on refining these advanced frameworks, for instance, by incorporating explicit peak-current constraints to directly manage stress or by developing hierarchical MPC structures to improve computational efficiency \cite{8799001}. Collectively, these advancements reveal a clear trajectory toward sophisticated multi-objective MPC designs, but they also highlight the dependence on increasingly complex analytical models under advanced modulation schemes.

This bespoke modeling paradigm ultimately leads to a series of obstacles. First, the modeling process is time-consuming and requires extensive expert effort, and its accuracy is difficult to verify. For instance, the analytical expression for the average inductor current in DAB converters under TPS modulation contains over 20 conditional branches \cite{10315245}, \cite{schwenzer2021review}. Second, parasitic parameters and dead-time effects that are often ignored to simplify computation continuously distort predicted results, leading to a discrepancy between theory and reality \cite{9525187}. Most critically, this approach completely lacks generality: any minor modification to the circuit topology or control objectives necessitates a complete re-derivation of the model and control law. This fragile one-off design process has become the primary barrier that prevents MPC from achieving widespread large-scale adoption.

In parallel with the analytical modeling challenges in the control domain, the rapid advancement of Digital Twin (DT) technology offers a highly promising alternative for power electronics modeling. DT technology emphasizes high fidelity and high automation, aiming to create virtual replicas that correspond precisely to physical entities. DT-based modeling frameworks, as demonstrated by both commercial real-time simulation platforms like RT-LAB and Typhoon HIL \cite{opalrt_rtlab, typhoon_hil, mpsoc} and numerous cutting-edge solutions emerging from academia \cite{10236507, 11077776, 9819434}, can automatically generate high-fidelity converter models directly from high-level descriptions such as a circuit netlist. By integrating numerical solvers and detailed device representations, DTs can intrinsically capture parasitic parameters, dead time, and other non-ideal behaviors, enabling an accurate reproduction of system dynamics \cite{10342664, 9819434}. This automated and unified modeling process significantly enhances both the fidelity and portability of models across different converter topologies, directly addressing the limitations of traditional analytical methods and showcasing the immense potential of DT technology \cite{zheng2025physicsembeddedneuralodessim2real, 11077776, zheng2025neuralsubstitutesolverefficient}.

In light of the inherent pain points of analytical MPC and capitalizing on the rapid progress in DT technology, this paper aims to propose a new direct synthesis paradigm to revolutionize the design workflow of power electronics controllers. To this end, this paper introduces a Digital-Twin-based Model Predictive Control (DT-MPC) framework that automatically generates executable control laws from a high-level system description. By completely replacing tedious manual equation derivation with a programmatically-generated high-fidelity DT model, the framework aims to transform control design from a complex mathematical exercise into an automated computational process, thereby paving a new path for developing universal, scalable, and high-performance power electronic controllers.

However, directly integrating a numerical DT into a real-time MPC presents three major computational challenges. First, the cost of real-time prediction using a DT model is formidable, due to intensive matrix computations and high-order numerical integrations \cite{10236507, 10342664}. Second, the high-fidelity DT provides a wealth of internal state variables, creating a complex objective management problem . Formulating a multi-objective cost function that robustly balances conflicting goals is non-trivial. Third, even with a well-defined cost function, the continuous control space of the converter creates a large non-convex optimization domain. Together with the infinite action set, it makes real-time optimization with traditional methods impractical \cite{8945348}. Discretizing the control space via a common compromise (e.g., gridding) is only effective for low-dimensional inputs and suffers from slow convergence.

To address these challenges, a DT-MPC framework is proposed, representing a paradigm shift in control design from manual model derivation to data-driven synthesis and automatic generation. Its main contributions include:

\begin{enumerate}
    \item \textbf{Direct MPC Controller Synthesis:} Direct conversion of circuit netlists into real-time high-performance controllers, eliminating manual derivation.
    
    \item \textbf{Physics-Informed Surrogate Model:} A neural predictor that accelerates high-fidelity digital twin calculations for high-frequency control.
    
    \item \textbf{Priority-based Online Optimization:} An efficient simplex-based optimizer with priority-based cost function to ensure convergence and balance of multi- objectives.
\end{enumerate}

The remainder of this paper is organized as follows. Section II defines the core problem of traditional MPC bottlenecks and introduces the fundamental concept of the DT-MPC framework. Section III details the complete methodology of the proposed DT-MPC, including the Neural Surrogate Predictor, the priority-based cost function, and the Simplex Search Optimizer. Section IV describes the practical deployment of the framework using a specific case study. Section V presents and analyzes the comprehensive experimental validation. Finally, Section VI concludes the paper.

\section{Problem Definition and DT-MPC Framework}
\label{sec:problem_and_framework}

This section first defines the fundamental challenges that the inherent hybrid-system nature of power electronic converters poses to traditional MPC. Subsequently, it introduces the DT-MPC framework as a paradigm shift and identifies the new computational challenges it brings.

\subsection{Problem Definition of MPC in Power Converters}
\label{sec:mpc_bottlenecks}

\begin{figure}[!t]
    \centering
    \includegraphics[width=0.49\textwidth]{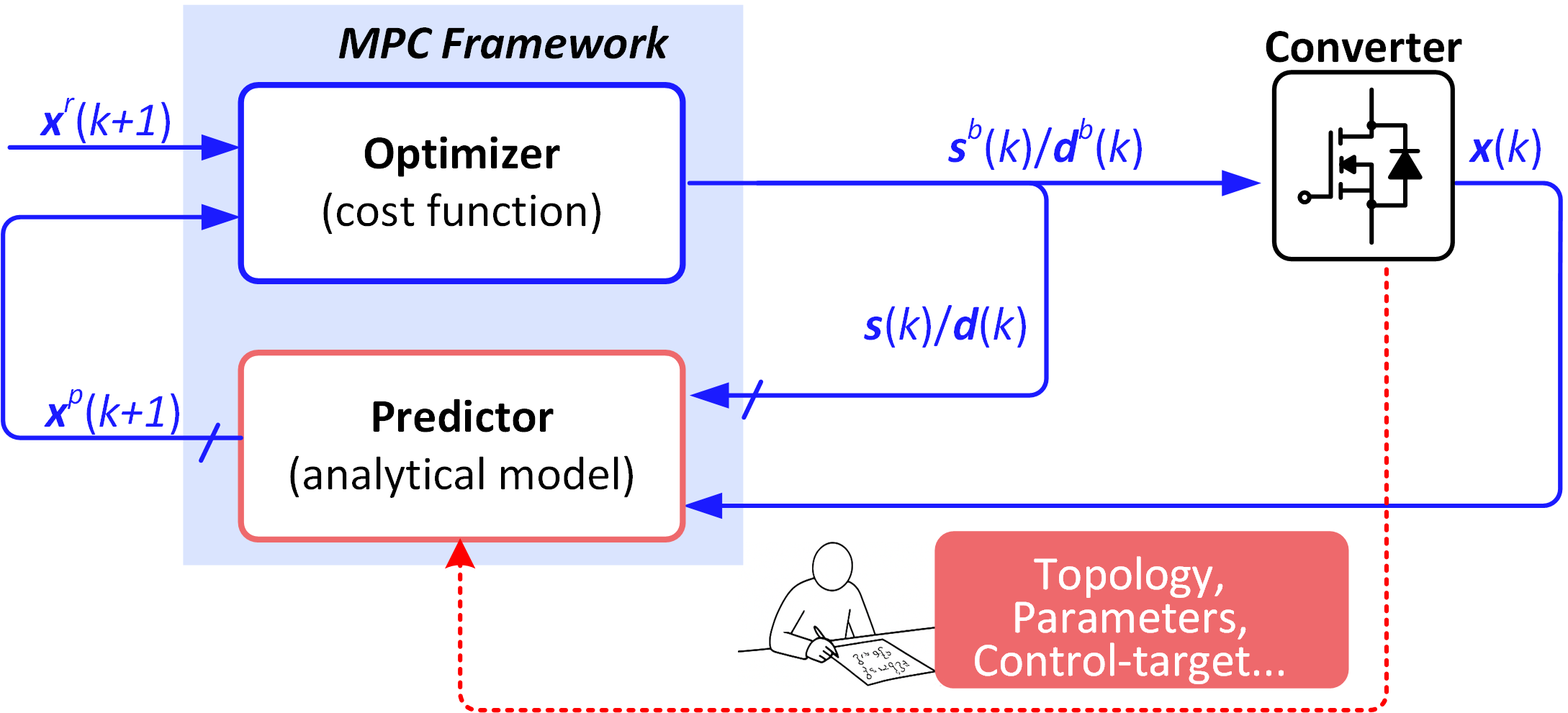} 
    \vspace{-0.8cm}
    \caption{The basic components of MPC}
    \label{fig:analytic_MPC}
\end{figure}

In MPC, the derivation of the control law depends on the system model's accurate prediction of future states. In its discrete-time form, an MPC is typically composed of a predictor and an optimizer, as is shown in Fig.~\ref{fig:analytic_MPC}. The predictor estimates the system's future dynamics, generally written as:
\begin{equation}
\label{eq:predictor}
\mathbf{x}_{k+1} = f(\mathbf{x}_k, \mathbf{u}_k)
\end{equation}

where $\mathbf{x}_k$ and $\mathbf{u_k}$ are the state vector (e.g., inductor currents and capacitor voltages) and the input vector of the system at the $k$-th step, and $f(\cdot)$ is the predictor function. The optimizer obatins an optimal control sequence with a cost function $J$ of the general form:
\begin{equation}
\label{eq:cost_function}
J = \sum_{i=0}^{N_p} \| \mathbf{y}_{k+i|k} - \mathbf{y}^*_{k+i} \|_Q^2 + \sum_{i=0}^{N_c-1} \|\mathbf{u}_{k+i|k}\|_R^2
\end{equation}
where $\mathbf{y}_k$ is the output vector at the $k^{th}$ step, and $N_p$ and $N_c$ are the prediction and control horizons, respectively. The control objective is to achieve reference tracking and constraint satisfaction by minimizing $J$ through receding horizon optimization.

\begin{figure*}[t]
\begin{minipage}{\textwidth}
\centering
\includegraphics[width=0.98\textwidth]{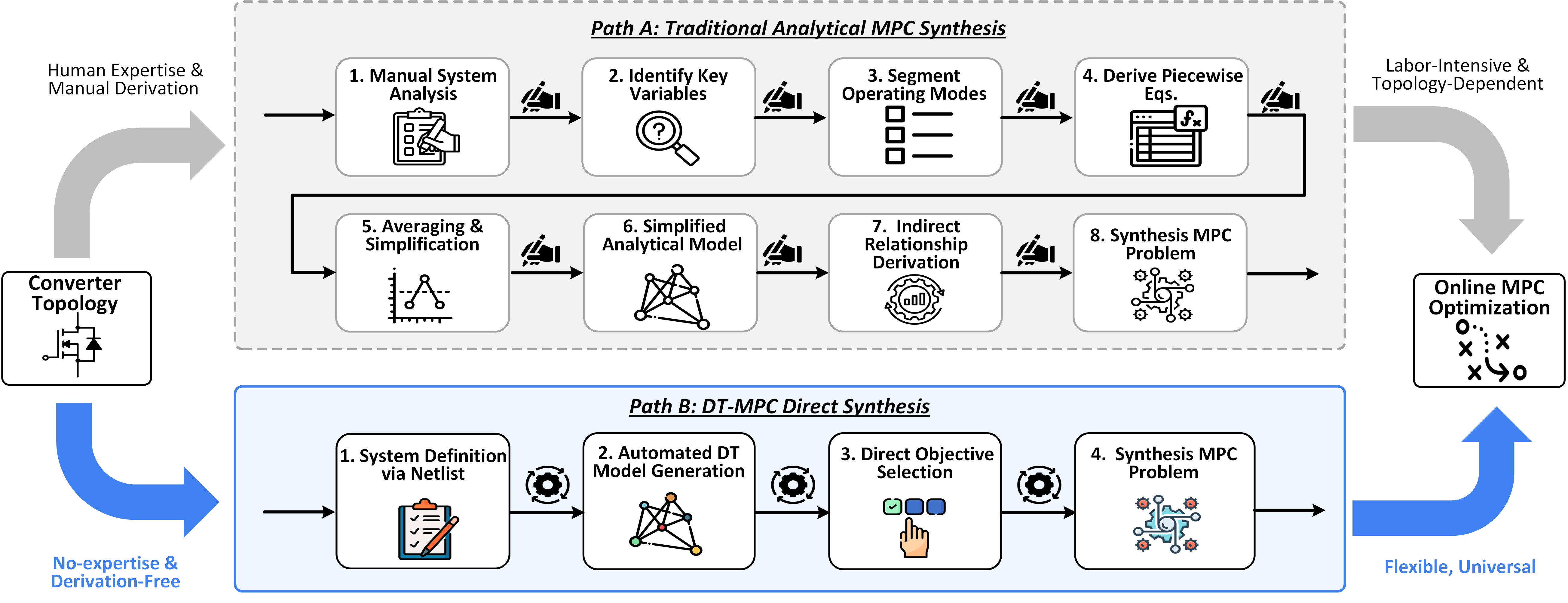}
\vspace{-0.3cm}
\caption{The proposed DT-MPC direct synthesis workflow.}
\label{fig:Direct_Synthesis}
\end{minipage}
\vspace{-0.4cm}
\end{figure*}

However, for power electronic converters that are inherently characterized by continuous physical dynamics and discrete switching events, $f(\cdot)$ is not in a fixed linear form. The dynamics can be described as a set of piecewise Ordinary Differential Equations (ODEs) based on its switching state $\mathbf{s}_k$, with the general state-space representation as:
\begin{equation}
\label{eq:state_space_piecewise}
\dot{\mathbf{x}}(t) = \mathbf{A}_{\mathbf{s}_k}\mathbf{x}(t) + \mathbf{B}_{\mathbf{s}_k}\mathbf{u}(t)
\end{equation}
where $\mathbf{A}_{\mathbf{s}_k}$ and $\mathbf{B}_{\mathbf{s}_k}$ are system matrices that change with $\mathbf{s}_k$. Consequently, $f(\cdot)$ is a complex hybrid nonlinear mapping controlled by  $\mathbf{s}_k$, making it difficult to obtain an analytical form.

To obtain a predictive model suitable for online optimization, traditional MPC typically resorts to simplified analytical models derived through averaging or piecewise linearization. The main processing steps are shown as Path A in Fig.~\ref{fig:Direct_Synthesis}. The purpose of this simplification is to avoid the high-frequency $n$-step numerical integration of ~\eqref{eq:state_space_piecewise} across a full control period. Instead, averaging methods approximate the system as a continuous smooth form, which yields a single-step prediction after discretization.
\begin{equation}
\label{eq:avg_model}
\dot{\mathbf{x}}(t) \approx \mathbf{A}_{\text{avg}}\mathbf{x}(t) + \mathbf{B}_{\text{avg}}\mathbf{u}(t) \implies \mathbf{x}_{k+1} = f_{\text{avg}}(\mathbf{x}_k, \mathbf{u}_k)
\end{equation}

In \eqref{eq:avg_model}, $f(\cdot)$ directly calculates $\mathbf{x}_{k+1}$ at the next control cycle from $\mathbf{x}_k$ at the current cycle.
All high-frequency intra-cycle dynamics are thus deliberately omitted. This manually-derived simplified model achieves computational tractability but introduces three insurmountable problems.

First, there are fidelity losses. The resulting simplified model from ~\eqref{eq:avg_model} is a coarse approximation of the true hybrid dynamics $f(\cdot)$ from ~\eqref{eq:state_space_piecewise}. It considers only partial output variables instead of the full $\mathbf{x}(t)$. Furthermore, the model is often tied to a specific control and modulation scheme (i.e., a specific $\mathbf{s}_k$), making it difficult to extend to other scenarios. More critically, key non-ideal effects such as dead time and high-frequency ripples are averaged out, leading to model mismatch and performance degradation.

Second, the simplification leads to objective limitations. Some critical performance metrics, such as ZVS conditions or transient peak currents, are instantaneous properties of $\mathbf{x}(t)$ and do not exist in the averaged model. It prevents engineers from directly including these goals when constructing the cost function. Consequently, they are forced to expend significant expert efforts (A.3-A.7 in Fig.~\ref{fig:Direct_Synthesis}) to find indirect relationships, attempting to optimize a hidden true objective by controlling an observable proxy variable. This approach lacks generality and fundamentally hinders the true co-optimization of multiple conflicting objectives.

Third, it results the optimizer to compromise for constraints. With a simplified $f(\cdot)$, the microsecond-level switching period of converters imposes extreme constraints on the optimizer for minimizing $J$. The most common solutions, such as Finite-Control-Set (FCS)-MPC or grid-search methods, are computationally feasible but coarse-grained. They perform a discretized search that cannot guarantee optimality in a continuous control space, and require extensive application-specific offline tuning to approximate a satisfactory solution.

\subsection{DT-MPC: The Direct Synthesis Paradigm}
\label{sec:dt_mpc_paradigm}

To overcome these limitations, this paper proposes a direct synthesis framework: DT-MPC, as shown in Path B of Fig.~\ref{fig:Direct_Synthesis}. The core premise of DT-MPC is to completely replace the manually-derived analytical model with a programmatically generated high-fidelity DT as a numerical predictor.

DTs are automatically synthesized from a high-level circuit netlist. The netlist is first abstracted into a directed graph and then converted into a set of state-space ODEs (i.e.,~\eqref{eq:state_space_piecewise}) that describe the fundamental dynamics for any switching state.

This new paradigm offers three transformative advantages: 1) The modeling process is algorithmically driven, eliminating time-consuming and error-prone manual derivation. 2) The DT model inherently captures all system dynamics, including non-ideal effects ignored by simplified models. 3) DT provides comprehensive observability into all internal states, enabling the optimizer to directly target any physical quantity by constructing a multi-objective $J$.

However, this ideal framework is not without cost. In essence, it shifts the engineering effort from \emph{offline manual derivation} to \emph{online computational load}. To be practical, DT-MPC must address three key challenges:

\begin{itemize}
\item \textbf{Computational speed:} Although the numerical DT provides high fidelity, it remains computationally expensive. The time required to solve its dynamics can barely accommodate the microsecond-level control cycle, leaving insufficient time for the optimizer to execute multiple iterations.
\item \textbf{Objective management:} The DT exposes a rich high-dimensional state space. A simple weighted-sum cost is inadequate, since secondary objectives (e.g., efficiency) may conflict with primary objectives (e.g., voltage regulation), degrading transient responses and inducing steady-state errors.

\item \textbf{Optimization complexity:} Combining a high-fidelity model with multi-objective optimization yields a high-dimensional and non-convex problem. Classical MPC optimizers (e.g., grid-based search) become prohibitive due to the curse of dimensionality.
\end{itemize}

\begin{figure*}[t]
\begin{minipage}{\textwidth}
\centering
\vspace{-0.5cm}
\includegraphics[width=0.95\textwidth]{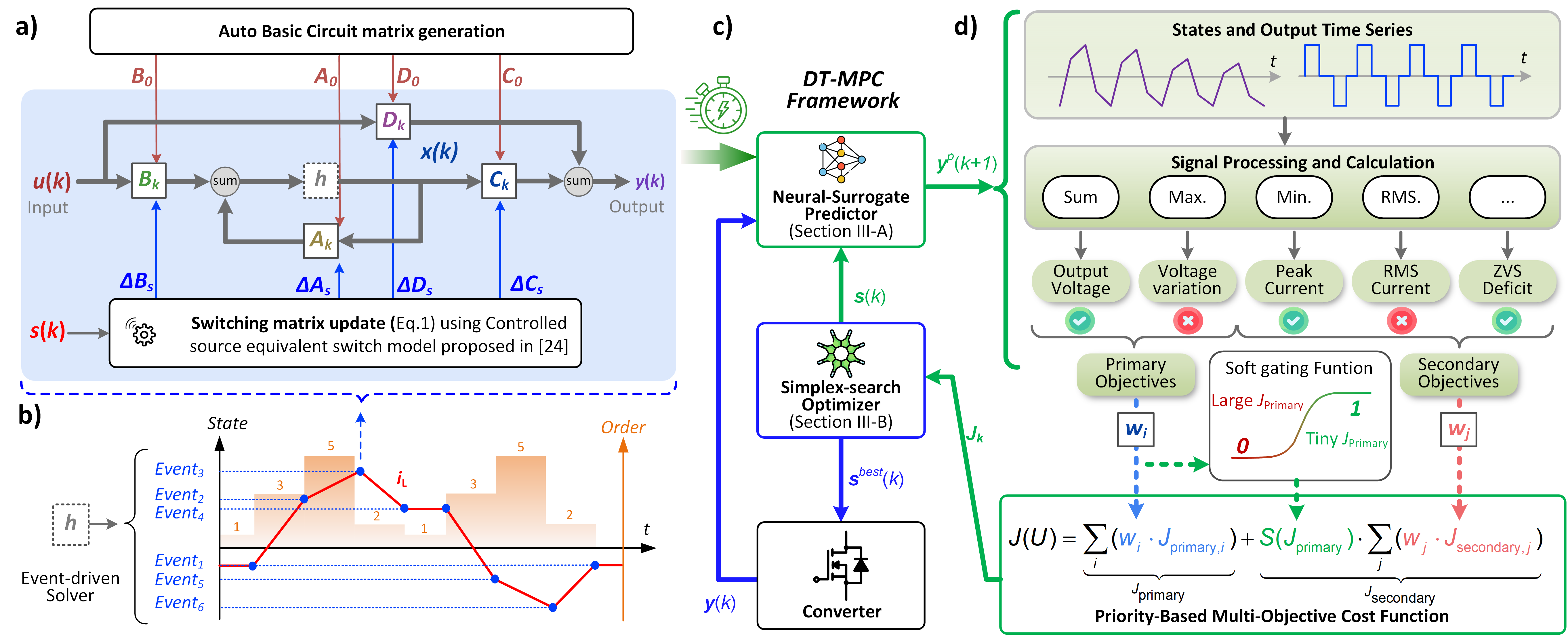}
\vspace{-0.3cm}
\caption{Core components of the DT-MPC framework. (a) Block diagram of the programmatically generated Digital Twin model. (b) Illustration of the event-driven solver. (c) DT-MPC framework. (d) The process of formulating the priority-based multi-objective cost function.}
\label{fig:dt_mpc_core_components}
\end{minipage}
\vspace{-0.4cm}
\end{figure*}

\section{Core Components and Implementation of DT-MPC}
\label{sec:core_components}

This section introduces the core components of DT-MPC designed to address the  aforementioned three challenges, including the Neural Surrogate Predictor (NSP), the priority-based cost function, and the Simplex Search Optimizer (SSO).

\subsection{Neural Surrogate Predictor for Real-Time Execution}
\label{sec:nsp}

The whole process of DT-MPC is shown in Fig.~\ref{fig:dt_mpc_core_components}. As established, directly using the numerical DT  as a predictor faces a severe computational speed challenge. This challenge has two layers: the online computation of system matrices and the online solving of model dynamics.

First, the system matrix computation adopting advanced switching models from \cite{10236507} is shown in Fig.~\ref{fig:dt_mpc_core_components}(a). Thus, $\mathbf{A}_{\mathbf{s}_k}$ in ~\eqref{eq:state_space_piecewise} can be expressed as: 
\begin{equation}
\mathbf{A}_{\mathbf{s}_k} = \mathbf{A}_{\mathbf{s}_0} + \Delta \mathbf{A}_{\mathbf{s}_k}
\end{equation}
where $\mathbf{A}_{\mathbf{s}_0}$ is the base matrix and $\Delta \mathbf{A}_{\mathbf{s}_k}$ is the increment matrix varying with $\mathbf{s}_k$. $\Delta \mathbf{A}(\mathbf{s}_k)$ can be further expressed as:
\begin{equation}
\label{eq:delta_A_full}
\Delta \mathbf{A}_{\mathbf{s}_k} = \mathbf{M}_{\mathbf{s}_k}\mathbf{K}_{\mathbf{s}_k}\mathbf{E}
\end{equation}
where $\mathbf{E}$ is a constant incidence matrix defined by the topology and $\mathbf{K}_{\mathbf{s}_k}$ is a diagonal matrix representing switch properties. Denoting $\mathbf{B}_s$ as the input matrix and $\mathbf{F}_s$ as a constant state-feedback matrix, the computation of $\mathbf{M}_{\mathbf{s}_k}$ involves a matrix inversion:
\begin{equation}
\label{eq:M_sk_inverse}
\mathbf{M}_{\mathbf{s}_k} = \mathbf{B}_s(\mathbf{I} - \mathbf{K}(\mathbf{s}_k)\mathbf{F}_s)^{-1}
\end{equation}
For a complex converter, ~\eqref{eq:M_sk_inverse} is an extremely computationally expensive step ($\mathcal{O}(m^3)$ where $m$ is the system dimension), making it unrealistic to perform online within a microsecond-level control cycle.

Second, for DTs using high-order event-driven solvers \cite{9819434} to accelerate processing while maintaining accuracy, as shown in Fig.~\ref{fig:dt_mpc_core_components}(b), it involves complex serial integration steps that are also time-consuming.

\begin{figure}[!t]
    \centering
    \includegraphics[width=0.49\textwidth]{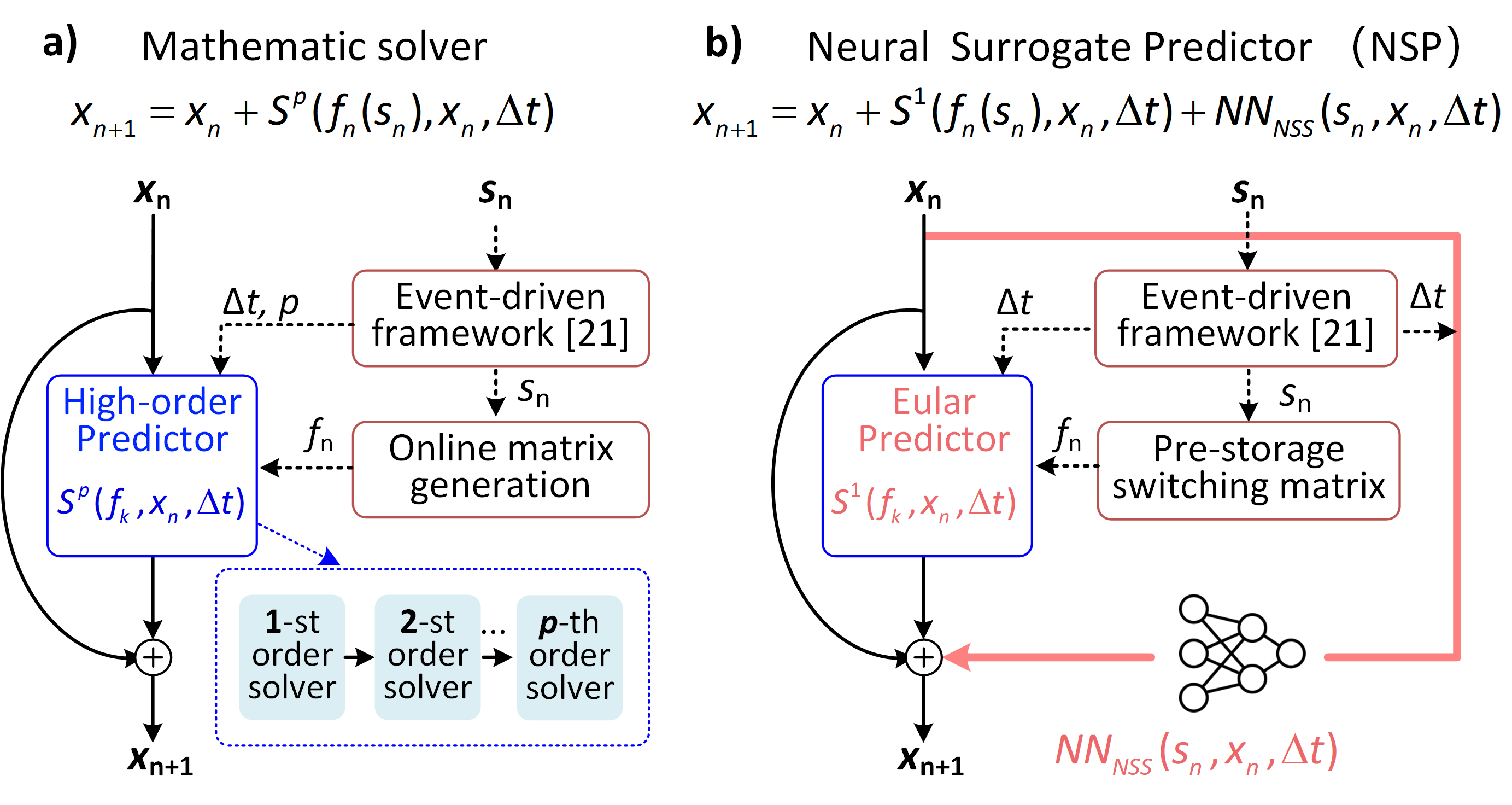} 
    \vspace{-0.8cm}
    \caption{(a) Traditional high-order mathematic solver. (b) Proposed neural surrogate predictor.}
    \label{fig:Neural Surrogate Predictor}
\end{figure}

To address both bottlenecks simultaneously, we proposed the NSP, as shown in Fig.~\ref{fig:Neural Surrogate Predictor}. The NSP is a physics-informed neural network (NN) designed to accelerate rather than replace the core physics.

To begin, we move the expensive matrix inversion in ~\eqref{eq:M_sk_inverse} entirely offline, so that all possible $\mathbf{M}_{\mathbf{s}_k}$ are pre-computed and stored in a Look-Up Table (LUT). Consequently, the online calculation of $\Delta \mathbf{A}_{\mathbf{s}_k}$ is reduced to a fast LUT access ($\mathcal{O}(1)$) and efficient matrix multiplications (~\eqref{eq:delta_A_full}).

Besides, the NSP does not learn the entire system dynamics from scratch. Instead, it utilizes a computationally efficient forward Euler solver as its physical core to provide a first-order state estimation:
\begin{equation}
\label{eq:forward_euler}
\mathbf{x}_{\text{FE}}(k+1) = \mathbf{x}(k) + h(\mathbf{A}_{\mathbf{s}_k} \mathbf{x}(k) + \mathbf{B}_s \mathbf{u}(k))
\end{equation}
where \(h\) is the integration time step, $\mathbf{x}(k)$, $\mathbf{u}(k)$ denote the value of the continuous-time $\mathbf{x}(t) $ and $\mathbf{u}(t) $ sampled at the \(k\)-th discrete time step, and the subcript FE represents the output of the forward Euler solver. Although this first-order approximation is fast, it suffers from truncation error. Therefore, we train a compact Forward Neural Network (FNN) $\mathbf{NN}_{\text{sur}}$ to learn and predict only the higher-order residual $R_k$, which is defined as the difference between the DT solution and the Euler solution:
\begin{equation}
\label{eq:residual_def}
R_k = \mathbf{x}_{\text{DT}}(k) - \mathbf{x}_{\text{FE}}(k)
\end{equation}
The final surrogate prediction $\mathbf{x}_{\text{sur}}$ is a high-fidelity combination of the physical solver and the data-driven correction term:
\begin{equation}
\label{eq:surrogate_final}
\mathbf{x}_{\text{sur}}(k+1) = \mathbf{x}_{\text{FE}}(k+1) + \mathbf{NN}_{\text{sur}}(\mathbf{x}(k), \mathbf{u}(k), \mathbf{s}_k, h; \boldsymbol{\theta})
\end{equation}
where $\boldsymbol{\theta}$ represents the network parameters.

In summary, the NSP transforms the high computational load of the DT into a parallelizable task by strategically combining two techniques: offline matrix computations and online physics-informed residual learning. This allows it to run at an extremely high speed, thereby securing the precious computational budget required for the subsequent optimizer.

\subsection{Priority-Based Cost Function for Multi-Objective Control}
\label{sec:cost_function}

The NSP can provide a high-fidelity full-state $\mathbf{x}_{\text{sur}}$. The next step is to leverage the state information to construct a proper cost function $J(\mathbf{U})$.

It is notable that the simple sum $J = \sum w_i J_i$ for different objectives $J_i$ with corresponding weights $w_i$ is not suitable for power converters. This is because priorities of $J_i$ are dynamic. For instance, during a large transient like a load step, the controller must prioritize the output voltage stability ($J_{\text{pri}}$) above all else. In steady state, however, the controller should optimize secondary objectives ($J_{\text{sec}}$) like ZVS or current stress, provided the stable voltage. Fixed $w_i$ cannot manage this conflict and may lead to secondary objectives interfering with the transient response and failing to optimize in steady state.

To resolve this, we proposed a priority-based cost function. As shown in Fig.~\ref{fig:dt_mpc_core_components}(d), the formulation distinguishes between $J_{\text{pri}}$ and a set of $J_{\text{sec},i}$. The core innovation is a soft-gating mechanism $S(J_{\text{pri}})$ (e.g., 1-$sigmoid(\cdot)$) which dynamically modulates the influence of $J_{\text{sec},i}$ based on the performance of $J_{\text{pri}}$:
\begin{equation}
\label{eq:priority_cost}
J(\mathbf{U}) = J_{\text{pri}} + S(J_{\text{pri}}) \cdot \sum_{i \in \text{sec}}(w_i \cdot J_{\text{sec},i})
\end{equation}

The mechanism follows a simple principle: During transients, when $J_{\text{pri}}$ is high, $S(J_{\text{pri}}) \approx 0$, which effectively blocks the secondary objectives, allowing the controller to focus on rapid stabilization and reference tracking. Conversely, in steady state, when the $J_{\text{pri}}$ is sufficiently low, $S(J_{\text{pri}}) \approx 1$. Then it activates the secondary objectives, permitting the optimizer to fine-tune performance by minimizing current stress or maximizing ZVS without compromising the primary goal. This priority-based formulation ensures a robust transient response without sacrificing steady-state optimality.

\subsection{Simplex Search Optimizer for  Control Space Exploration}
\label{sec:sso}

To efficiently optimize the high-dimensional, non-convex, and computationally expensive cost function $J(\mathbf{U})$, we employ an efficient, derivative-free Simplex Search Optimizer (SSO).

Traditional brute-force methods like grid search are infeasible, as their computational cost grows exponentially with the control dimension $n$ (e.g., $3^n - 1$ evaluations). With each evaluation requiring a call to the NSP, this exponential complexity is unacceptable under real-time constraints.

\begin{figure}[!t]
    \centering
    \includegraphics[width=0.49\textwidth]{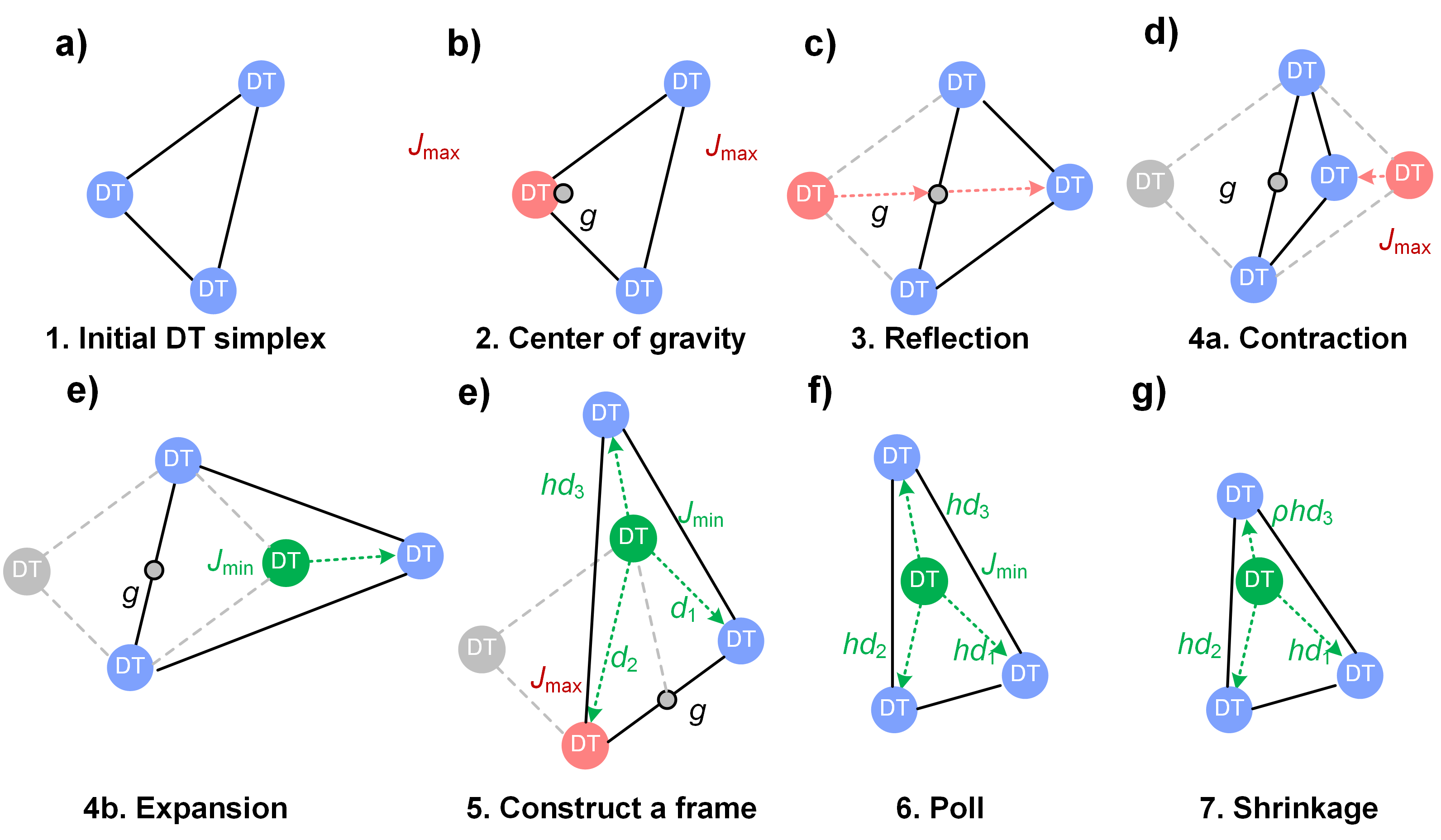} 
    \vspace{-0.8cm}
    \caption{Geometric steps of one simplex search iteration. (a) Initial simplex. (b) Identify worst vertex ($J_{max}$) and centroid ($g$). (c) Reflect $J_{max}$ through $g$. (d) Expand simplex if reflection is very good. (e) contract if reflection is poor. (f) If no improvement, poll directions around best vertex ($J_{min}$). (g) If polling fails, shrink simplex towards $J_{min}$.}
    \label{fig: Simplex Search}
\end{figure}

In contrast, the SSO maintains a geometric figure called a simplex, which consists of $n+1$ vertices in an $n$-dimensional control space, as illustrated in Fig.~\ref{fig: Simplex Search},. Each vertex $\mathbf{U}_i$ represents a complete control input vector (e.g., $\mathbf{U}_i = [D_0, D_1, D_2]$ for triple-phase-shift modulation).  The SSO algorithm iteratively replaces the vertex with the worst cost ($J_{\text{max}}$) by finding a new, better point through a series of geometric transformations: Reflection, Expansion, and Contraction.

A key feature of this algorithm is its convergence guarantee. If the standard transformations fail to find a better point, the algorithm initiates a Positive Basis Search   from the current best vertex ($\mathbf{U}_0$) and refines the search in a smaller, more promising region by   Shrinking   the simplex (convergence proof provided in Appendix A).

In detail, the current simplex can be set as $\mathcal{S} = \{U_0, \dots, U_n\}$, with its vertices sorted according to their costs, such that:
\begin{equation}
    J(U_0) \leq J(U_1) \leq \dots \leq J(U_n)
\end{equation}
At each iteration, the simplex search aims to find a new vertex $\mathbf{U}_{\text{new}}$ that satisfies a sufficient decrease condition:
\begin{equation}
\label{eq:sso_decrease}
J(\mathbf{U}_{\text{new}}) \le J(\mathbf{U}_0) - \sigma h_s^2
\end{equation}
where $h_s$ is the search scale and $\sigma$ is a positive constant. To find such a point, the algorithm first find the centroid $\mathbf{U}_c$ of all vertices except the worst one, $\mathbf{U}_n$:
\begin{equation}
\label{eq:sso_centroid}
\mathbf{U}_c = \frac{1}{n}\sum_{i=0}^{n-1}\mathbf{U}_i
\end{equation}
Then, the algorithm performs geometric transformations based on the centroid $\mathbf{U}_c$ ,
\begin{itemize}
    \item \textbf{Expansion:} If $U_r$ yields the lowest cost (i.e., $J(U_r) < J(U_0)$), the algorithm attempts an expansion to $U_e = U_c + \gamma (U_r - U_c)$. The better of $U_e$ and $U_r$ is accepted.
    \item \textbf{Reflection:} If the cost of $U_r$ is better than the second-worst vertex (i.e., $J(U_0) \leq J(U_r) < J(U_{n-1})$), then $U_r$ is accepted.
    \item \textbf{Contraction:} If the reflection point $U_r$ is poor (i.e., $J(U_r) \geq J(U_{n-1})$), the algorithm performs a contraction, either outside ($U_{\text{oc}}$) or inside ($U_{\text{ic}}$), to find a point closer to the simplex's interior.
\end{itemize}

If standard transformations fail, the search proceeds from $\mathbf{U}_0$ along the positive basis directions $\mathcal{V}^+$:
\begin{equation}
\label{eq:sso_basis_search}
\mathbf{U}_0 + h_s \mathbf{v}_j, \quad j=1, ..., n+1
\end{equation}
If all search directions fail, the search scale is reduced:
\begin{equation}
\label{eq:sso_shrink}
h_s \leftarrow \rho h_s, \quad \text{where } 0 < \rho < 1
\end{equation}
Most critically, the computational cost of SSO scales  linearly  ($n+1$) with the control dimension $n$, not exponentially. This superior efficiency allows the optimizer to perform multiple iterations within a single control cycle (in the time secured by the NSP), enabling rapid convergence to the optimal solution. The whole algorithm of SSO is shown in Algorithm~\ref{alg:compact_simplex_v2}

\begin{algorithm}[ht]
\caption{Compact NSP-Accelerated SSO}
\label{alg:compact_simplex_v2}
\begin{algorithmic}[1]
\Require Initial simplex $\mathcal{S}=\{U_0, \dots, U_n\}$, scale $h$, constants $\sigma, \rho$
\State Sort $\mathcal{S}$ so that $J(U_0) \leq \dots \leq J(U_n)$
\State $U_c \leftarrow \frac{1}{n} \sum_{i=0}^{n-1} U_i$
\State $U_r \leftarrow U_c + (U_c - U_n)$ \Comment{Calculate reflection point}
\State \Comment{Determine candidate point $U_{cand}$ from transformations}
\If{$J(U_r) < J(U_0)$}
    \State $U_e \leftarrow U_c + 2(U_r - U_c)$
    \State $U_{cand} \leftarrow \arg\min_{U \in \{U_r, U_e\}} J(U)$ \Comment{Expansion}
\ElsIf{$J(U_r) < J(U_{n-1})$}
    \State $U_{cand} \leftarrow U_r$ \Comment{Accept Reflection}
\Else
    \State $U_{cand} \leftarrow \begin{cases} U_c + 0.5(U_r - U_c) & \text{if } J(U_r) < J(U_n) \\ U_c - 0.5(U_c - U_n) & \text{otherwise} \end{cases}$ \Comment{Contraction}
\EndIf
\State \Comment{Check candidate; if it fails, search from best point}
\If{$J(U_{cand}) \leq J(U_0) - \sigma h^2$}
    \State $U_n \leftarrow U_{cand}$ \Comment{Accept candidate}
\Else
    \State $found\_better \leftarrow \textit{false}$
    \For{each basis direction $v_j \in \mathcal{V}^+$}
        \If{$J(U_0 + h v_j) \leq J(U_0) - \sigma h^2$}
            \State $U_n \leftarrow U_0 + h v_j$; $found\_better \leftarrow \textit{true}$; \textbf{break}
        \EndIf
    \EndFor
    \If{\textbf{not} $found\_better$}
        \State $h \leftarrow \rho h$; Re-initialize simplex around $U_0$. \Comment{Shrink}
    \EndIf
\EndIf
\end{algorithmic}
\end{algorithm}

\subsection{Overall Implementation Workflow}
\label{sec:workflow}

\begin{figure}[!t]
    \centering
    \includegraphics[width=0.49\textwidth]{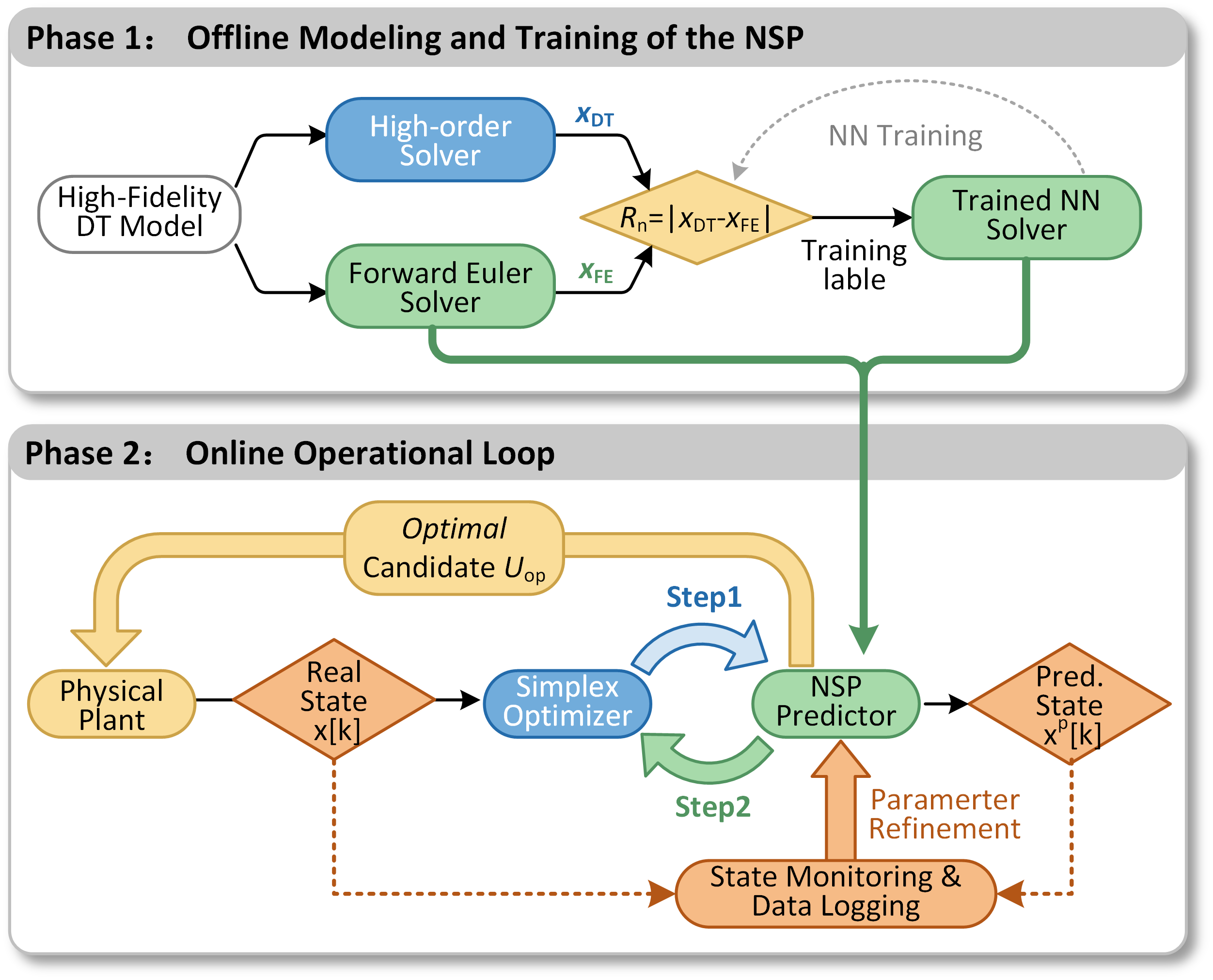} 
    \vspace{-0.8cm}
    \caption{The two-phase methodological workflow of DT-MPC. 
    }
    \label{fig: workflow of DT-MPC}
\end{figure}

  Finally, we integrate the NSP, the priority-based cost function, and the SSO into a cohesive two-phase workflow  , as shown in Fig.~\ref{fig: workflow of DT-MPC}.

  Phase 1 is Offline Training:   This phase is executed once before deployment. The high-fidelity DT model is simulated across a wide range of operating conditions to generate a comprehensive dataset. This dataset is then used to train the $NN_{\text{sur}}$ (the core of the NSP) by minimizing the residual loss function:
\begin{equation}
\label{eq:loss_residual}
\mathcal{L}_{\text{residual}} = \frac{1}{M}\sum_{n=0}^{M-1} \|\mathbf{R}_n - NN_{\text{sur}}(\mathbf{x}(n), \mathbf{u}(n), h, \mathbf{s}_{k})\|_2^2
\end{equation}

  Phase 2 is the Online Operational Loop:   This loop executes within every control cycle. First, the SSO proposes a set of candidate control inputs $\{\mathbf{U}_i\}$ (the vertices of its current simplex), typically using the optimal solution from the previous cycle as a warm start. Subsequently, for each candidate $\mathbf{U}_i$, the hardware-accelerated NSP is invoked to perform an ultra-fast prediction, and its corresponding priority-based cost $J(\mathbf{U}_i)$ is immediately calculated. Based on the resulting set of costs $\{J(\mathbf{U}_i)\}$, the SSO updates its simplex (via reflection, contraction, etc.) to find a better candidate. This "NSP-predict, SSO-update" iteration repeats for a fixed number of times within the control period budget. Finally, the vertex with the lowest cost, $\mathbf{U}_{\text{optimal}}$, is applied to the physical converter for the next switching cycle.

\section{ Case Setup and DT-MPC Deployment}

\begin{figure*}[t]
\begin{minipage}{\textwidth}
\centering
\vspace{-0.5cm}
\includegraphics[width=1.0\textwidth]{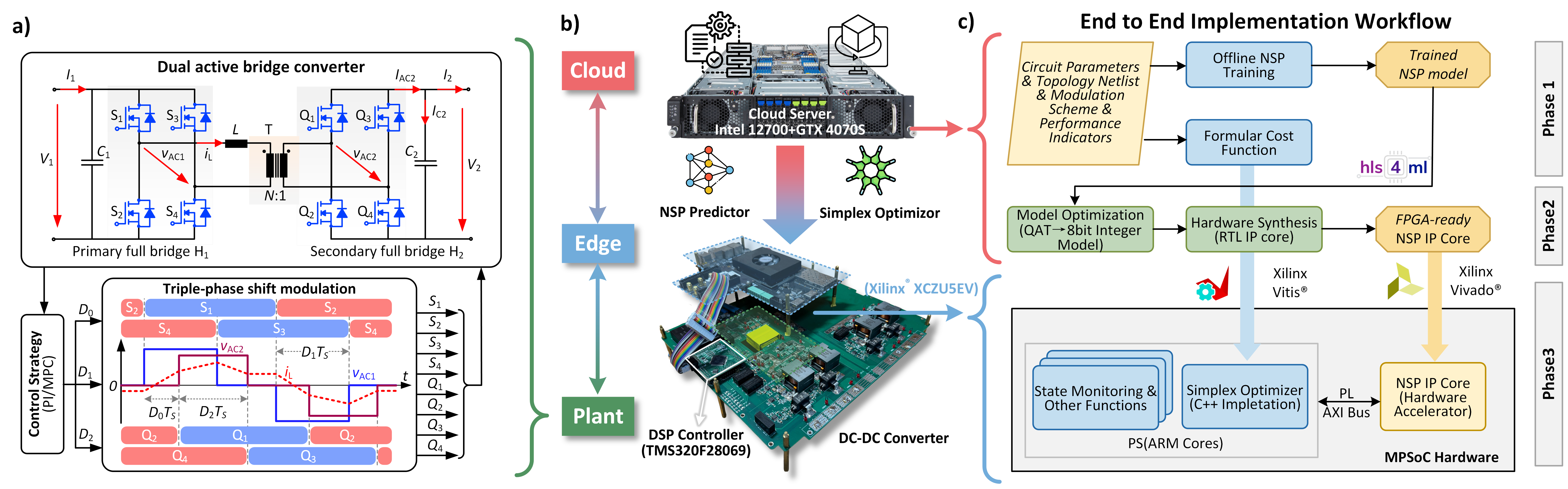}
\vspace{-0.8cm}
\caption{Overall System Framework and Workflow. (a) DAB converter topology with TPS modulation. (b) Cloud-Edge-Plant three-tier control architecture. (c) End-to-end implementation workflow from offline training to online deployment.}
    \label{fig:overall_workflow_short}
\end{minipage}
\vspace{-0.4cm}
\end{figure*}

This section introduces a case study of high-frequency DAB converters to validate the proposed DT-MPC framework through a comprehensive case study. First, the DAB converter is introduced and and justified as a challenging benchmark. Then the complete experimental system architecture is detailed. Finally, an end-to-end implementation workflow of the DT-MPC framework is applied to the DAB converter.

\subsection{Case Study: DAB Converter}
The DAB converter is a canonical high-frequency topology that provides bidirectional power flow and galvanic isolation. As shown in Fig.~\ref{fig:overall_workflow_short}(a), it consists of two H-bridges, a high-frequency transformer (turns ratio  $N:1$), and a leakage inductor $L$. Under TPS modulation, the converter is controlled by three independent phase-shift ratios ($D_0, D_1, D_2$), offering extra degrees of freedom for multi-objective optimization, such as efficiency improvement by achieving ZVS or minimizing current stress.

The DAB converter serves as an ideal and challenging benchmark for validating the DT-MPC framework for several reasons. First, its high switching frequency imposes stringent real-time constraints (usually microseconds) on the control algorithm. Second, TPS modulation creates a three-dimensional continuous control space, which makes the optimization computationally expensive for conventional methods. Finally, its complex non-linear dynamics, including ZVS conditions and peak inductor current behaviors, are notoriously difficult to model accurately with analytical equations. These challenges perfectly highlight the necessity for DT-MPC.

\subsection{System Architecture: Cloud, Edge, and Plant}
\label{ssec:platform}

An experimental platform is constructed to validate the effectiveness of the proposed DT-MPC, as illustrated in Fig.~\ref{fig:overall_workflow_short}(b), which comprises three interconnected parts:
\begin{itemize}
    \item \textbf{Cloud Platform:} An NVIDIA GeForce 4070 Super GPU and Intel Core i7-12700 CPU server are used for the offline training of the NSP.
    \item \textbf{Edge Platform:} A Xilinx Zynq UltraScale+ MPSoC (XCZU5EV) board serves as the real-time controller. This heterogeneous platform integrates a multi-core Arm-based Processing System (PS) and Programmable Logic (PL) and is ideal for hardware-software co-design.
    \item \textbf{Physical Plant:} The key parameters of a 1500\,W 100\,kHz DAB converter prototype are listed in Table~\ref{tab:params}. The PWM generation and protection are handled by a Texas Instrument TMS320F28069 DSP.
\end{itemize}

For interconnections, the edge platform communicates with the physical plant through a 20\,MHz SPI bus and connects to the cloud platform through Ethernet for data transfer and model updates.

\subsection{End-to-End Implementation Workflow}

The practical realization of the DT-MPC algorithm for the DAB converter follows a systematic end-to-end workflow. This workflow connects high-level system configuration with real-time edge deployment in three distinct phases, as illustrated in Fig.~\ref{fig:overall_workflow_short}(c).

\subsubsection{Phase 1: Framework Configuration and Offline Training}
The workflow begins with the circuit netlist, which defines the converter topology and component parameters. From this, the performance objectives are formulated to include DC output voltage regulation, peak inductor current (related to conduction losses), and ZVS operation (related to switching losses). Different modulation schemes (1-D SPS, 2-D DPS, and 3-D TPS) are configured, corresponding to SSOs with 2, 3, and 4 vertices, respectively. Subsequently, the NSP is trained following the procedure outlined in Section~III-C. The key hyperparameters used for training, such as the final MLP architecture and learning rate, are summarized in Table~\ref{tab:nsp_params}.

\subsubsection{Phase 2: Model Preparation and Hardware Integration}
Once trained, the NSP model is prepared for efficient hardware implementation. The PyTorch model undergoes Quantization-Aware Training (QAT) and structured pruning, converting its parameters and activations into an 8-bit integer format suitable for embedded deployment. This quantized model is then translated into a synthesizable RTL IP core using the HLS4ML toolchain. During this process, customized matrix-vector threshold units are employed to exploit the intrinsic parallelism of the neural network computations and maximize throughput on the FPGA fabric.

\subsubsection{Phase 3: Edge Deployment and Collaborative Design}
The final runtime deployment utilizes a partitioned, collaborative architecture to execute the operational loop.
\begin{itemize}
    \item \textbf{SSO implemented on PS:} The sequential, decision-based logic of the optimizer is implemented in C++ and executed on the flexible ARM cores of the Processing System (PS).
    \item \textbf{NSP implemented on PL:} The computationally intensive NSP is deployed as a dedicated hardware accelerator IP core within the Programmable Logic (PL) of the FPGA.
\end{itemize}
This architecture forms a tightly coupled operational loop. The PS manages the overall optimization process, offloading prediction requests for each candidate solution to the PL-based NSP accelerator via the AXI bus. The PL also receives real-time ADC-sampled state data from the physical converter, allowing it to perform the state-monitoring function by comparing predictions with measurements. After the optimization converges, the final optimal phase-shift ratios are transmitted from the PS to an external DSP to generate the PWM signals for the next switching cycle. This integrated co-design ensures that the DT-MPC operates efficiently within the stringent temporal constraints of high-frequency converters.

\begin{table}[!t]
\renewcommand{\arraystretch}{1.3}
\vspace{-10pt}
\caption{Parameters of the DAB Converter}
\label{tab:params}
\vspace{-8pt}
\centering
\begin{tabular*}{\columnwidth}{@{\extracolsep{\fill}} l l l l}
\toprule
\textbf{Parameter} & \textbf{Value} & \textbf{Parameter} & \textbf{Value} \\
\midrule
Rated Power & 1500 W & Switching Freq. ($f_s$) & 100 kHz \\
Input Voltage ($V_{\text{in}}$) & 48 V & Transformer Ratio & 1:1 \\
Output Voltage Range & 16-32 V & Line Inductor ($L$) & 1.13 $\mu$H \\
Rated Output Voltage ($V_{\text{out}}$) & 24 V & Capacitors ($C_1, C_2$) & 230 $\mu$F \\
\bottomrule
\end{tabular*}
\end{table}

\begin{table}[!t]
\renewcommand{\arraystretch}{1.3}
\vspace{-10pt}
\caption{NSP Architecture, Training, and Deployment Parameters}
\vspace{-8pt}
\label{tab:nsp_params}
\centering
\begin{tabular*}{\columnwidth}{@{\extracolsep{\fill}} l l l l}
\toprule
\multicolumn{4}{c}{\textbf{NSP Architecture \& Training}} \\
\midrule
Network Architecture & MLP  & Learning Rate & $1 \times 10^{-3}$ \\
Activation Function & ReLU & Training Epochs & 100 \\
Loss Function & MSE & Batch Size & 256 \\
Training Optimizer & Adam & Network Size & \{4,32,32,3\}\\
\midrule
\multicolumn{4}{c}{\textbf{NSP Hardware Deployment}} \\
\midrule
Deployment Toolchain & HLS4ML & Pruning Method & Magnitude \\
Quantization Method & QAT & Weight Bits & 8 (Integer) \\
Target Hardware & PL side & Output Precision & FP16 \\
\bottomrule
\end{tabular*}
\end{table}

\section{Experimental Results and Performance Evaluation}

This section validates the DT-MPC framework through a series of quantitative evaluations and experiments. It first assesses the accuracy and speed of the NSP. Then it demonstrates the superiority of the SSO over conventional gradient-free methods. Finally, it presents experiments and compares the DT-MPC against various benchmark controllers.

\subsection{Performance Evaluation of NSP}

\begin{figure*}[t]
\begin{minipage}{\textwidth}
\centering
\vspace{-0.5cm}
\includegraphics[width=1\textwidth]{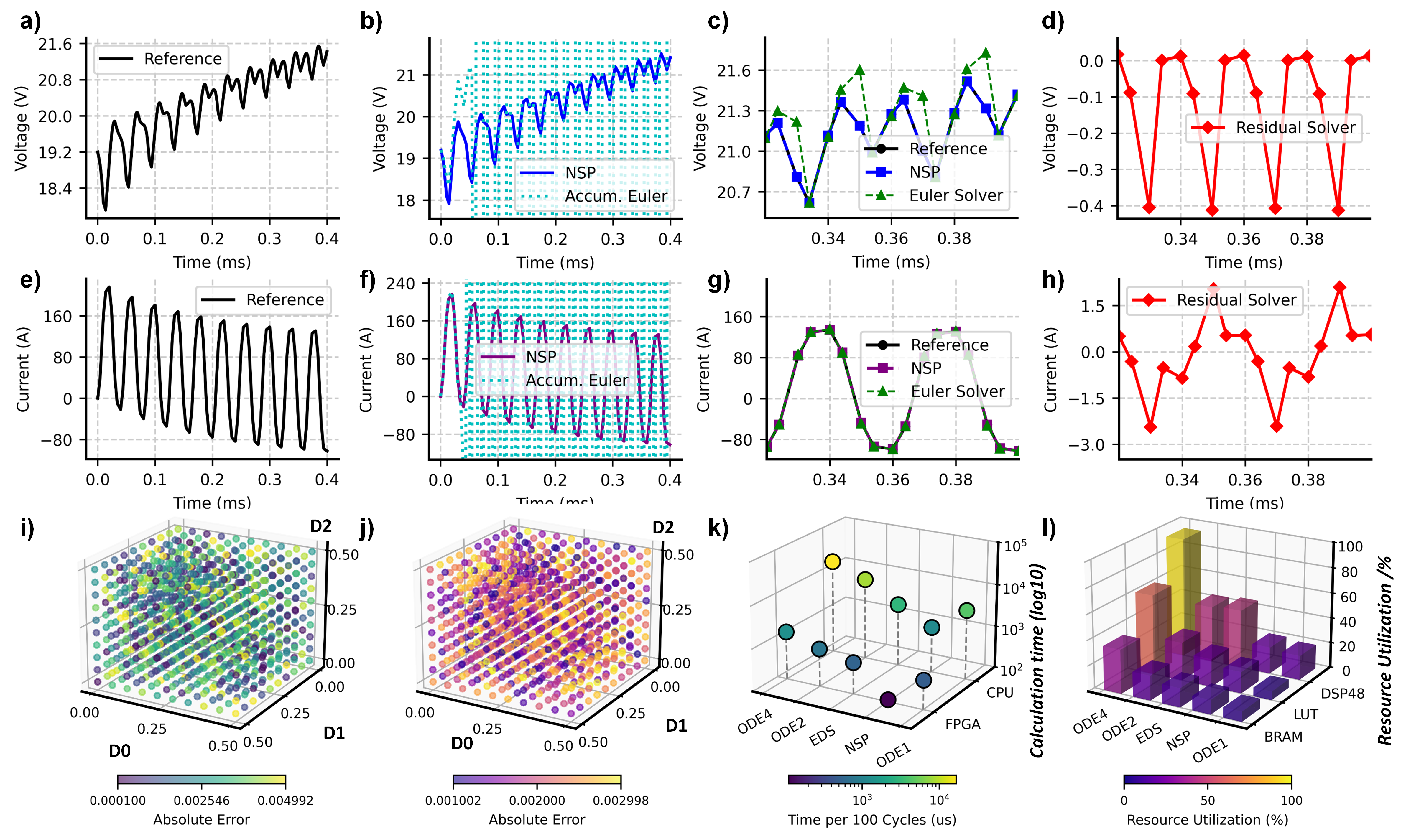}
\vspace{-0.7cm}
\caption{Performance evaluation of the NSP. 
    (a), (e) High-fidelity reference waveforms for output voltage and inductor current. 
    (b), (f) Long-term prediction comparison, showing the divergence of the pure Euler solver. 
    (c), (g) Zoomed-in view of transient accuracy, comparing the NSP against the reference and the Euler solver. 
    (d), (h) Corrective output from the residual-learning network. 
    (i), (j) Absolute error distribution of the NSP across the TPS control space for (i) output voltage and (j) inductor current. 
    (k), (l) Performance benchmark against other numerical solvers, detailing (k) computation speed and (l) FPGA resource utilization.}
    \label{fig:NSP_performance_evaluation}
\end{minipage}
\vspace{-0.4cm}
\end{figure*}

\begin{table*}[!t]
\renewcommand{\arraystretch}{1.5}
\caption{Performance Comparison of Different Numerical Predictors}
\vspace{-5pt}
\label{tab:solver_performance_comparison}
\centering
\begin{tabular}{l c c c c c c c c}
\toprule
\textbf{Solver} & \multicolumn{2}{c}{\textbf{Points \& Step Size Analysis}} & \multicolumn{2}{c}{\textbf{Ave. Execution Time\textsuperscript{a,b}}} & \textbf{Speedup\textsuperscript{c}} & \multicolumn{3}{c}{\textbf{FPGA Resource Utilization (\%)}} \\
\cmidrule(lr){2-3} \cmidrule(lr){4-5} \cmidrule(lr){6-6} \cmidrule(lr){7-9}
& \textbf{Per Control Cycle} & \textbf{Step Size} & \textbf{CPU (ms)} & \textbf{FPGA ($\mu$s/cycle)} & \textbf{(CPU / FPGA)} & \textbf{BRAM} & \textbf{LUT} & \textbf{DSP48} \\
\midrule
Dopri5 (Reference) & 22--34 & Variable & 48.4 & N/A & 3.42x / N/A & N/A & N/A & N/A \\
ODE4 & 67 (Step = 150 ns) & Fixed & 165.4 & 13.6 & 1.00x / 12.13x & 34.7 & 62.9 & 94.6 \\
ODE2 & 134 (Step = 75 ns) & Fixed & 85.0 & 7.8 & 1.95x / 21.21x & 19.5 & 30.0 & 44.2 \\
EDS & 10--16 & Variable& 29.6 & 5.48 & 5.59x / 30.18x & 15.0 & 22.0 & 46.0 \\
ODE1 & 200 (Step = 50ns) & Fixed  & 45.2& 4.88 & 3.66x/ 33.89x & 7.86 & 6.79 & 19.2 \\
\textbf{NSP (Proposed)} & \textbf{8} & \textbf{Variable} & \textbf{13.6 } & \textbf{1.12} & \textbf{12.16x  /147.68x} & \textbf{11.3} & \textbf{16.9} & \textbf{21.1} \\
\bottomrule
\multicolumn{9}{p{\dimexpr\textwidth-2\tabcolsep}}{\textsuperscript{a}\footnotesize CPU platform: Intel i7-12700 (single core), 32GB RAM. \textsuperscript{b}FPGA platform: Xilinx XCZU5EV (Total PL resources: 256K LUTs, 1248 DSPs, 26.6 Mb BRAM). \textsuperscript{c}Speedup is calculated relative to the ODE4 solver on the CPU platform.}
\end{tabular}
\vspace{-0.6cm}
\end{table*}

\begin{figure*}[t]
\begin{minipage}{\textwidth}
\centering
\vspace{-0.5cm}
\includegraphics[width=0.97\textwidth]{HyperMPC_Exp02.png}
\vspace{-0.5cm}
\caption{Optimization trajectories of different gradient-free search methods. 
(a) 1-D optimization under SPS control. 
(b) 2-D optimization with cost function contours under DPS control. 
(c) 3-D optimization in the TPS control space. 
For all cases, the initial starting point is (0.25, 0.25, 0.25). 
(a1)-(c1) Trajectory for grid search. 
(a2)-(c2) Trajectory for adaptive grid search. 
(a3)-(c3) Trajectory for the proposed simplex search.}
\label{fig:optimizer_trajectories}
\end{minipage}
\vspace{-0.3cm}
\end{figure*}

\begin{figure*}[t]
\begin{minipage}{\textwidth}
\centering
\includegraphics[width=0.97\textwidth]{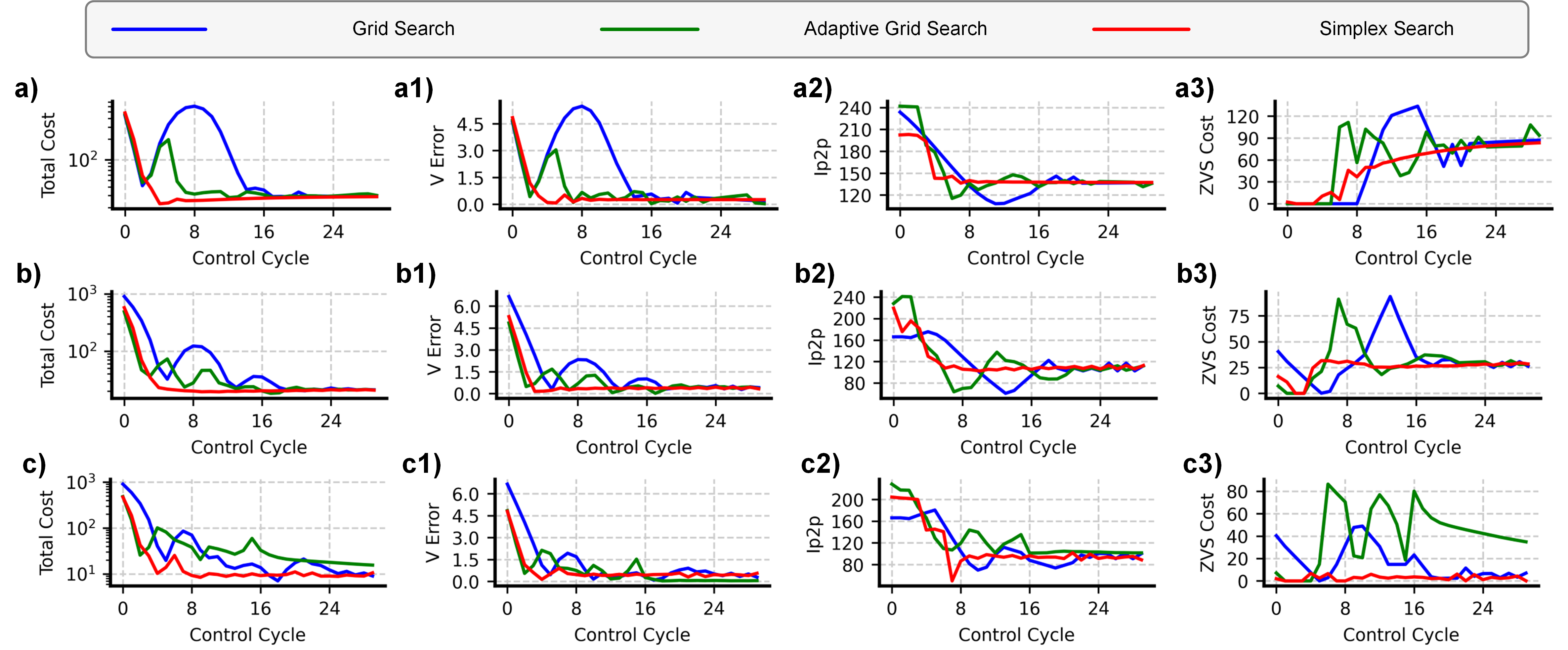}
\vspace{-0.4cm}
\caption{Cost function convergence for different optimizers under various control strategies. 
(a) Total cost of SPS, (a1) voltage tracking error, (a2) peak-to-peak inductor current, and (a3) ZVS deficit under SPS control. 
(b) Total cost of DPS, (b1) voltage tracking error, (b2) peak-to-peak inductor current, and (b3) ZVS deficit under DPS control. 
(c) Total cost of TPS, (c1) voltage tracking error, (c2) peak-to-peak inductor current, and (c3) ZVS deficit under TPS control. 
The ZVS deficit quantifies the current required to meet the ZVS condition.}
\label{fig:cost_convergence}
\end{minipage}
\vspace{-0.6cm}
\end{figure*}

The computational accuracy and speed of the NSP are evaluated and presented in Fig.~\ref{fig:NSP_performance_evaluation} and Table~\ref{tab:solver_performance_comparison}. The results of a high-precision Dopri5 solver are used as the ground-truth reference, and the pure Forward Euler method is used as a baseline to highlight the higher-order residual learning. Note that the case utilizes an open-loop DPS control mode to ensure that performance differences among the solvers originate solely from their respective algorithms.

As shown in Figs.~\ref{fig:NSP_performance_evaluation}(a) and~\ref{fig:NSP_performance_evaluation}(b), the accumulated Euler solver diverges over time due to the continuous accumulation of its large truncation error, while the output voltage and inductor current under NSP exhibit high fidelity with the reference. The zoomed-in views shown in Figs.~\ref{fig:NSP_performance_evaluation}(c) and ~\ref{fig:NSP_performance_evaluation}(d) reveal the accuracy mechanism. For the NSP, the initial value for each step is the corrected solution from the previous step because the corrective output from the residual-learning network (i.e., Figs.~\ref{fig:NSP_performance_evaluation}(e) and ~\ref{fig:NSP_performance_evaluation}(f)) precisely compensates for the error of the base Euler solver at each step. To quantify this precision, the NSP is evaluated across the entire TPS control space. The absolute error distributions across thousands of operating points for the output voltage in Fig.~\ref{fig:NSP_performance_evaluation}(g) and the inductor current in Fig.~\ref{fig:NSP_performance_evaluation}(h) have verified its high precision with maximum errors being negligible. Furthermore, the key metrics of the NSP and other solvers are visualized in Figs.~\ref{fig:NSP_performance_evaluation}(i) and~\ref{fig:NSP_performance_evaluation}(j), which indicates that the NSP demonstrates exceptional speed and efficiency while maintaining high accuracy. 

The overall performance comparison of the NSP against other numerical solvers is concluded in Table~\ref{tab:solver_performance_comparison}. Among the solvers, NSP uses variable step sizes and compute only at the switching moments, resulting in a small number of computational points (i.e., 8 points) per control cycle. It is significantly fewer than fixed-step methods like ODE4 (67 points) and ODE2 (134 points) that must use sub-microsecond steps to maintain accuracy. The improved computational efficiency also translates to a superior speed. In terms of CPU, the execution time of NSP is 13.6\,ms, which is over 12 times faster than the baseline ODE4 solver. This advantage is even magnified on the FPGA side, where NSP achieves an execution time of 1.12\,$\mu$s per cycle, resulting in a remarkable speedup of over 147 times when compared to the CPU baseline. For the 100\,kHz DAB converter (i.e., 10\,$\mu$s period), the FPGA-based NSP can enable 9 iterations per control cycle. It is noteworthy that the high performance is achieved with a modest utilization of hardware resources as shown in Fig.~\ref{fig:NSP_performance_evaluation}(j), which verifies that NSP is a fast and resource-efficient predictor with significant faster-than-real-time capabilities.

\subsection{Performance Evaluation of SSO}

\begin{figure*}[t]
\begin{minipage}{\textwidth}
\centering
\vspace{-0.4cm}
\includegraphics[width=1\textwidth]{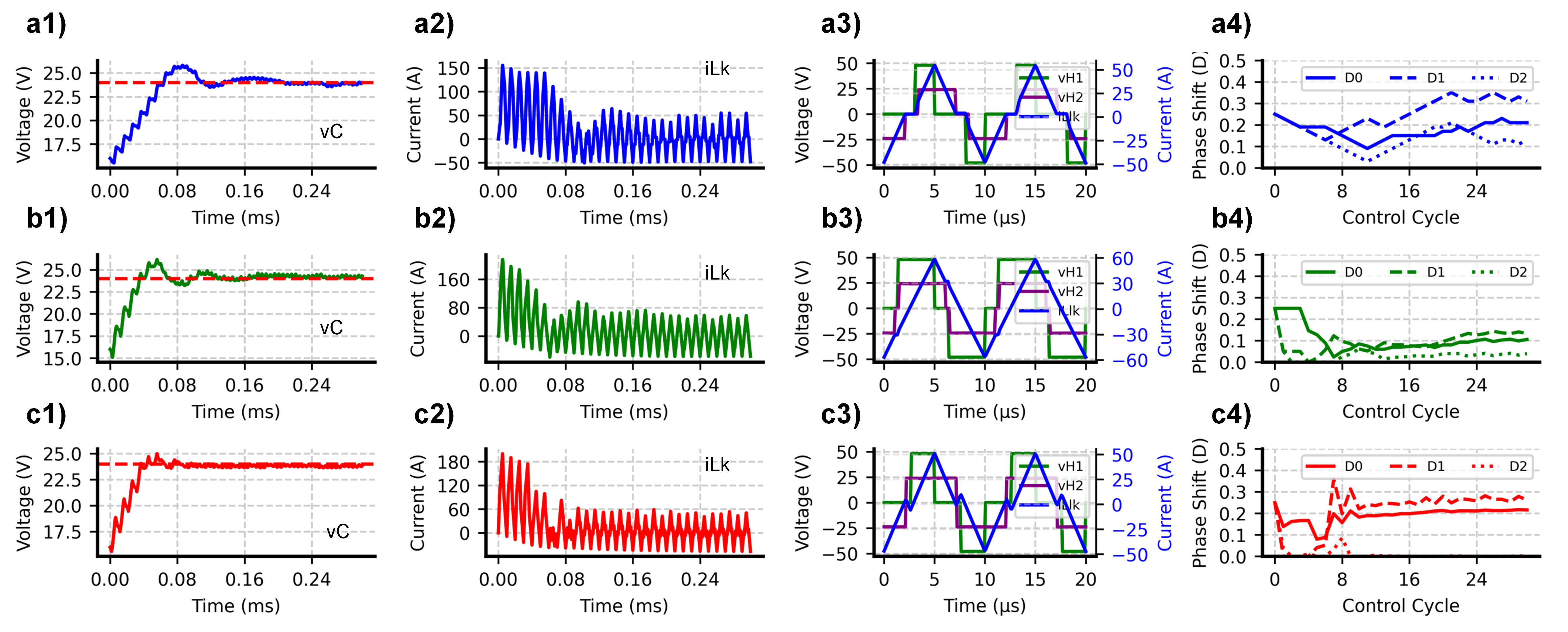}
\vspace{-0.9cm}
\caption{Dynamic performance of different optimizers under TPS control. The rows compare the performance of (a1-a4) Grid Search, (b1-b4) Adaptive Grid Search, and (c1-c4) the proposed Simplex Search. For each method, the columns show: (1) the output voltage ($v_C$) transient response, (2) inductor current ($i_{L}$), (3) H-bridge voltages for ZVS visualization, and (4) the evolution of the phase-shift ratios ($D_0, D_1, D_2$).}
\label{fig:tps_optimizer_results}
\end{minipage}
\vspace{-0.3cm}
\end{figure*}

The superiority of the proposed SSO is evaluated against commonly used gradient-free methods (e.g., grid search and adaptive grid search). All three optimizers use the same trained NSP as their predictor to optimize a composite objective of voltage tracking, peak-to-peak current minimization, and ZVS operation under different control dimensions (i.e., SPS, DPS, and TPS). A comprehensive comparison is summarized in Table~\ref{tab:optimizer_comparison}, with dynamic results shown in Figs.~\ref{fig:optimizer_trajectories}, ~\ref{fig:cost_convergence}, and ~\ref{fig:tps_optimizer_results}.

\begin{table*}[!t]
\renewcommand{\arraystretch}{1.3}
\caption{ Quantitative Comparison of Gradient-Free Optimizers}
\label{tab:optimizer_comparison}
\vspace{-0.2cm}
\centering
\begin{tabular*}{\textwidth}{@{\extracolsep{\fill}} l l l l}
\toprule
\textbf{Feature} & \textbf{Grid Search} & \textbf{Adaptive Grid Search} & \textbf{Simplex Search (Proposed)} \\
\midrule
\textbf{Evaluation Points (per Iteration)} & Exponential ($3^n-1$), 26 for TPS & Exponential ($3^n-1$), 26 for TPS & Linear ($n+1$) \\
\textbf{Search Flexibility} & Rigid (fixed axes) & Rigid (fixed axes) & Flexible (unrestricted) \\
\textbf{Convergence Behavior} & Slow ($\approx$30 cycles) & Moderate ($\approx$20 cycles) & Fast ($\approx$10 cycles) \\
\textbf{Dynamical Performance (TPS Case)} 
& Sensitive to step size 
&  Prone to oscillations 
& Smooth and stable operation \\
\textbf{Overall Robustness} 
& Low
& Moderate 
& High \\
\bottomrule
\end{tabular*}
\end{table*}

First of all, the fundamental characteristics of the optimizers are analyzed, as detailed in Table~\ref{tab:optimizer_comparison}. demonstrates remarkable computational efficiency, as it requires only $n+1$ evaluation point per iteration, exhibiting a linear growth with respect to the control dimension  $n$. In contrast, the evaluation points in grid-based methods grow exponentially (i.e., $3^n-1$), which leads to a computational explosion. For example, 26 evaluation points are required per iteration for TPS under grid search. Meanwhile, it offers superior search flexibility due to its unrestricted geometric transformations instead of the rigid axis-aligned grid search.

Therefore, the SSO has more efficient optimization trajectories, as visualized in Fig.~\ref{fig:optimizer_trajectories}. Although starting from the same initial point, the SSO consistently finds a more direct path to the optimal solution, especially in DPS (Fig.~\ref{fig:optimizer_trajectories}(b)) and TPS (Fig.~\ref{fig:optimizer_trajectories}(c)) spaces.

The impact of the computational efficiency of the optimizers on convergence is quantified in Fig.~\ref{fig:cost_convergence}. As the control dimension increases from SPS (row a) to TPS (row c), the SSO consistently achieves the fastest convergence and the lowest total cost, whereas the grid search exhibits the slowest convergence and the poorest dynamic response. Although the grid search can eventually reach results comparable to those of the simplex method with sufficiently small step sizes, its performance degrades severely with larger dimensions. The adaptive grid search, which involves multiple interdependent parameters such as base step size, gain, and threshold, converges faster but also exposes the fundamental limitation of grid-based methods in escaping local optima. Consequently, it often becomes trapped in locally optimal regions when dealing with multi-objective coordination, as shown in Fig.~\ref{fig:cost_convergence}(c3). The large initial cost value further drives it to adopt oversized step lengths, leading to oscillations and premature convergence. In contrast, the SSO optimizer, equipped with a positively oriented basis search, maintains global search capability and provides guaranteed convergence even in high-dimensional spaces.

Finally, Fig.~\ref{fig:tps_optimizer_results} illustrates the closed-loop performance in the most challenging TPS case. The SSO demonstrates the best overall performance, achieving a fast and smooth dynamic response as shwon inFig.~\ref{fig:tps_optimizer_results}(c1), the lowest peak-to-peak current as shown in Fig.~\ref{fig:tps_optimizer_results}(c2), and superior ZVS operation as shown in Fig.~\ref{fig:tps_optimizer_results}(c3). Due to its superior computaional efficiency, the SSO quickly settles on an optimal set of phase-shift ratios, as shown in Fig.~\ref{fig:tps_optimizer_results}(c4), outperforming the more oscillatory and sub-optimal responses produced by the grid-search methods.

\subsection{Experimental Comparison}

\begin{figure*}[t]
\begin{minipage}{\textwidth}
\centering
\includegraphics[width=1\textwidth]{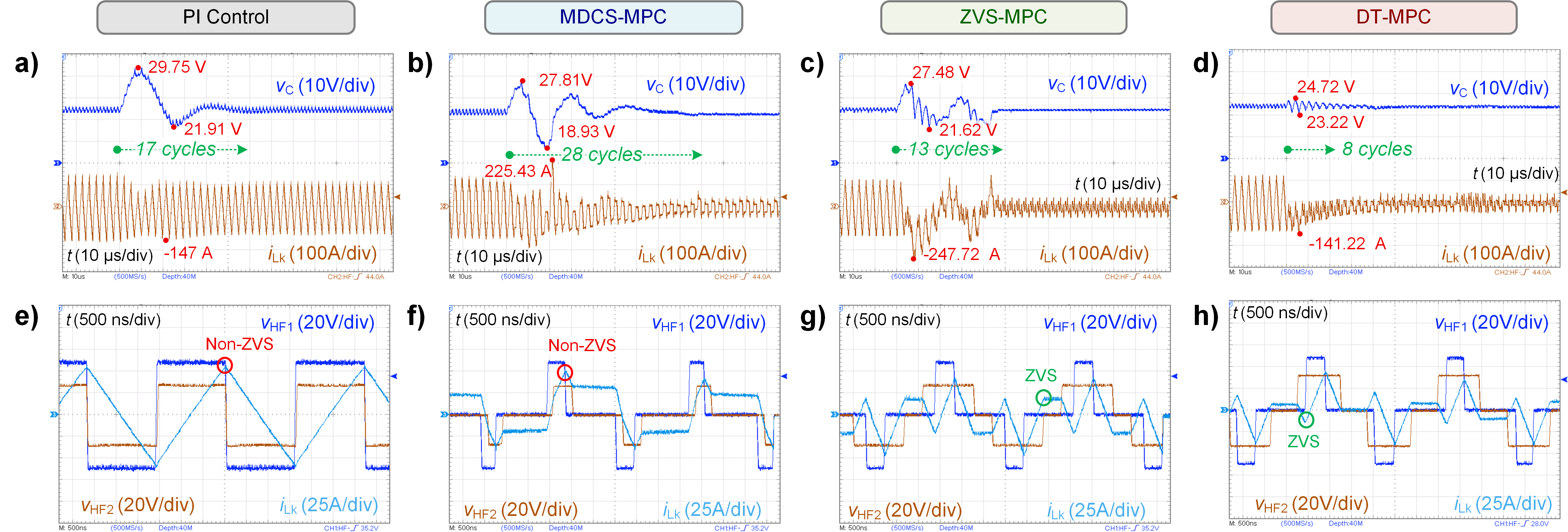}
\vspace{-0.7cm}
\caption{Experimental results for a load step transient from 100\% to 10\% rated load with $V_{ref} = 24 V$. (a)-(d) Dynamic response of the output voltage ($v_C$) and inductor current ($i_{L}$) for PI, MDCS-MPC, ZVS-MPC, and the proposed DT-MPC, respectively. (e)-(h) Zoomed-in steady-state waveforms of the high-frequency voltages and currents for each controller.}
\label{fig:load_step_results}
\end{minipage}
\end{figure*}

\begin{figure*}[t]
\begin{minipage}{\textwidth}
\centering
\vspace{-0.3cm}
\includegraphics[width=1\textwidth]{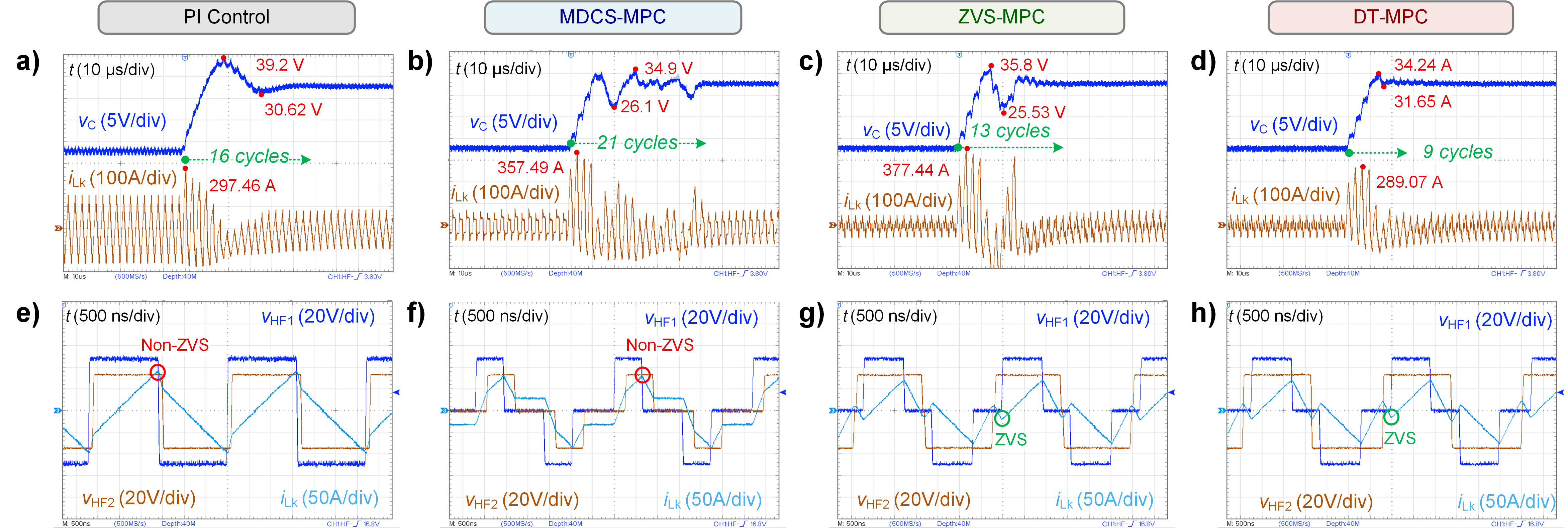}
\vspace{-0.7cm}
\caption{Experimental results for a reference voltage step transient from 16\,V to 32\,V with 30\% rated load. (a)-(d) Dynamic response of the output voltage ($v_C$) and inductor current ($i_{L}$) for PI, MDCS-MPC, ZVS-MPC, and the proposed DT-MPC, respectively. (e)-(h) Zoomed-in steady-state waveforms of the high-frequency voltages and currents for each controller after settling.}
 \label{fig:voltage_step_results}
\end{minipage}
\vspace{-0.5 cm}
\end{figure*}

This section evaluates the complete performance of DT-MPC against a conventional PI controller and two state-of-the-art analytical MPC strategies (i.e., MDCS-MPC \cite{8799001} and ZVS-MPC \cite{10315245}). All these three MPC are implemented under the TPS modulation framework. Specifically, the MDCS-MPC evaluates 26 candidate points per switching period and minimizes the peak-to-peak inductor current based on pre-derived analytical current expressions obtained from offline analysis. In contrast, the ZVS-MPC strategy regulates the output voltage through MPC control of $D_0$, while $D_1$ and $D_2$ are computed from pre-derived steady-state equations. It should be noted that both strategies are inherently constrained by their dependence on offline derivations. All control strategies share unified control objectives-achieving fast output voltage transitions while minimizing the high-frequency current ripple and maximizing ZVS operation. The detailed comparison among these controllers is summarized in Table~\ref{tab:controller_comparison}.

For all the strategies, experiments with a large load step and a large reference voltage step have been conducted to comprehensively evaluate their dynamic and steady-state performances.

The dynamic performances of the four controllers under a large load step (100\% to 10\% load) and a large voltage reference step (16\,V to 32\,V) are shown in Fig.~\ref{fig:load_step_results} and Fig.~\ref{fig:voltage_step_results}, respectively. The quantitative performance metrics are detailed in Table~\ref{tab:controller_comparison}. Across both test cases, the proposed DT-MPC demonstrates a significantly superior dynamic response. 

As shown in Figs.~\ref{fig:load_step_results}(d) and~\ref{fig:voltage_step_results}(d), the DT-MPC settles in just 8-9 cycles with the smallest voltage deviation (1.5\,V and 2.59\,V). This is a largely because the optimizer is flexible to find the optimal combination of phase shifts during the transient to rapidly regulate the output voltage. In contrast, the ZVS-MPC is constrained by its steady-state derived equations and operates with only one effective DOF for voltage regulation during the transient, leading to a slow and highly oscillatory response (13 cycles in the load step), as shown in Figs.~\ref{fig:load_step_results}(c) and~\ref{fig:voltage_step_results}(c). The MDCS-MPC shows more significant oscillations and a longer settling time (28 cycles in the load step), which reveals the inefficiency of grid-based discrete optimization when applied to high-dimensional switching decisions under large disturbances. The PI controller that lacks multi-objective capabilities exhibits the largest overshoot and slowest response.

The DT-MPC also achieves excellent steady-state performance, as shown in Figs.~\ref{fig:load_step_results}(h) and~\ref{fig:voltage_step_results}(h). The DT-MPC consistently maintains full ZVS while PI and MDCS-MPC exhibit non/ partial-ZVS features. Furthermore, as detailed in Table~\ref{tab:controller_comparison}, the peak-to-peak inductor current ($i_{Lpp}$) of the DT-MPC is comparable to that of the MDCS-MPC, which is specifically designed for current minimization.

\begin{figure*}[t]
\begin{minipage}{\textwidth}
\centering
\vspace{-0.3cm}
\includegraphics[width=1\textwidth]{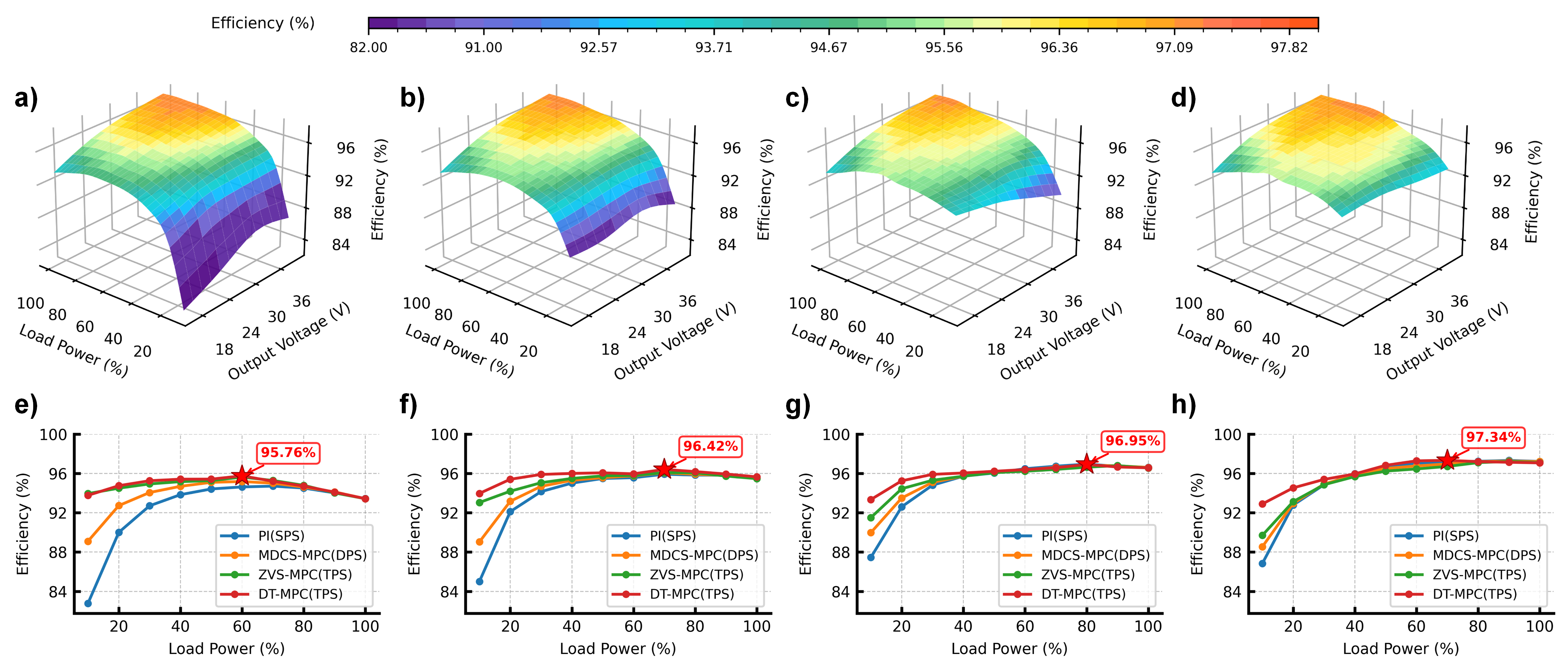}
\vspace{-0.7cm}
\caption{Steady-state efficiency comparison of different controllers. (a)-(d) Efficiency maps over a wide range of output voltages and load powers for (a) PI, (b) MDCS-MPC, (c) ZVS-MPC, and (d) the proposed DT-MPC. (e)-(h) Efficiency curves versus load power at fixed output voltages of (e) 16\,V, (f) 24\,V, (g) 32\,V, and (h) 40\,V, highlighting the peak efficiency achieved by each controller.}
\label{fig:efficiency_comparison}
\end{minipage}
\vspace{-0.5cm}
\end{figure*}

The comprehensive efficiency comparison presented in Fig.~\ref{fig:efficiency_comparison} provides the final validation of the proposed method. The efficiency maps (Figs.~\ref{fig:efficiency_comparison}(a)-(d)) and cross-sectional curves (Figs.~\ref{fig:efficiency_comparison}(e)-(h)) illustrate that the DT-MPC consistently operates at or near the highest efficiency points over a wide operational range. Notably, the DT-MPC demonstrates superior performance even at light loads, surpassing the specialized ZVS-MPC. This result warrants a deeper explanation, as the primary benefit of DT-MPC is typically considered to be its dynamic response. The reason for this enhanced steady-state efficiency lies in the fundamental difference between the underlying optimization models. The ZVS-MPC relies on an analytical model that is optimized exclusively for achieving Zero-Voltage Switching, assuming ZVS is the dominant factor for high efficiency. However, this single-objective approach can be suboptimal, particularly at light loads where maintaining full ZVS may require significant circulating currents, thereby elevating conduction losses ($I^2 R$).

In contrast, the DT-MPC leverages its high-fidelity DT, which captures a more complete profile, including peak-to-peak values (related with conduction losses), ZVS deficit(related with switching losses) and others. Its multi-objective cost function can therefore perform a true online optimization. Consequently, the DT-MPC autonomously discovers the true optimal operating point. The efficiency maps therefore validate that this holistic optimization approach not only delivers superior dynamic performance but also achieves the highest steady-state efficiency.

\begin{table*}[!t]
\renewcommand{\arraystretch}{1.4}
\caption{Comprehensive Comparison of Different Controllers}
\vspace{-0.2cm}
\label{tab:controller_comparison}
\centering
\footnotesize
\begin{tabular*}{\textwidth}{@{\extracolsep{\fill}} l l l l c c c c c}
\toprule
\textbf{Controller} & \textbf{Modeling Basis} & \textbf{Dynamic DOFs} & \textbf{Multi-Obj.} & \textbf{Settling Cycles$^{\textrm{a}}$} & \textbf{Volt. Dev.$^{\textrm{a}}$ (V)} & \textbf{Max. $i_{L}$$^{\textrm{a}}$ (A)} & \textbf{ZVS}  & \textbf{$i_{Lpp}$ (A)}\\
\midrule
PI Control & N/A & 1 & Low &  17 /  16 & 7.84 / 8.58 & 147 / 297 & 4/8 & 242 / 190.14 \\
MDCS-MPC \cite{8799001} & Analytical  & 2(Fixed) & Medium & 28 /21 &  8.88/ 8.77 & 225.43 / 357.49 & 6/8 & 101 / 164.61\\
ZVS-MPC \cite{10315245} & Analytical  & 1 (Effective) & Medium & 13 / 13 & 5.86/ 10.34 & 242.72 / 37.447 & 8/8 & 90.02 / 148.35 \\
DT-MPC  & Numerical & 3 (Full) & High &  8 / 9 & 1.5 / 2.59 & 141.22 / 289.07 & 8/8 & 81.14 / 147.35 \\
\bottomrule
\end{tabular*}
\begin{minipage}{\textwidth}
\vspace{0.1cm}
\footnotesize
$^{\textrm{a}}$Performance metrics are presented for Case 1 / Case 2 respectively. Case 1: Load Step (100\% $\to$ 10\%). Case 2: Voltage Step (16V $\to$ 32V).
\end{minipage}
\vspace{-0.7cm}
\end{table*}

\subsection{Comprehensive Discussion}

\begin{figure}[!t]
    \centering
    \includegraphics[width=0.49\textwidth]{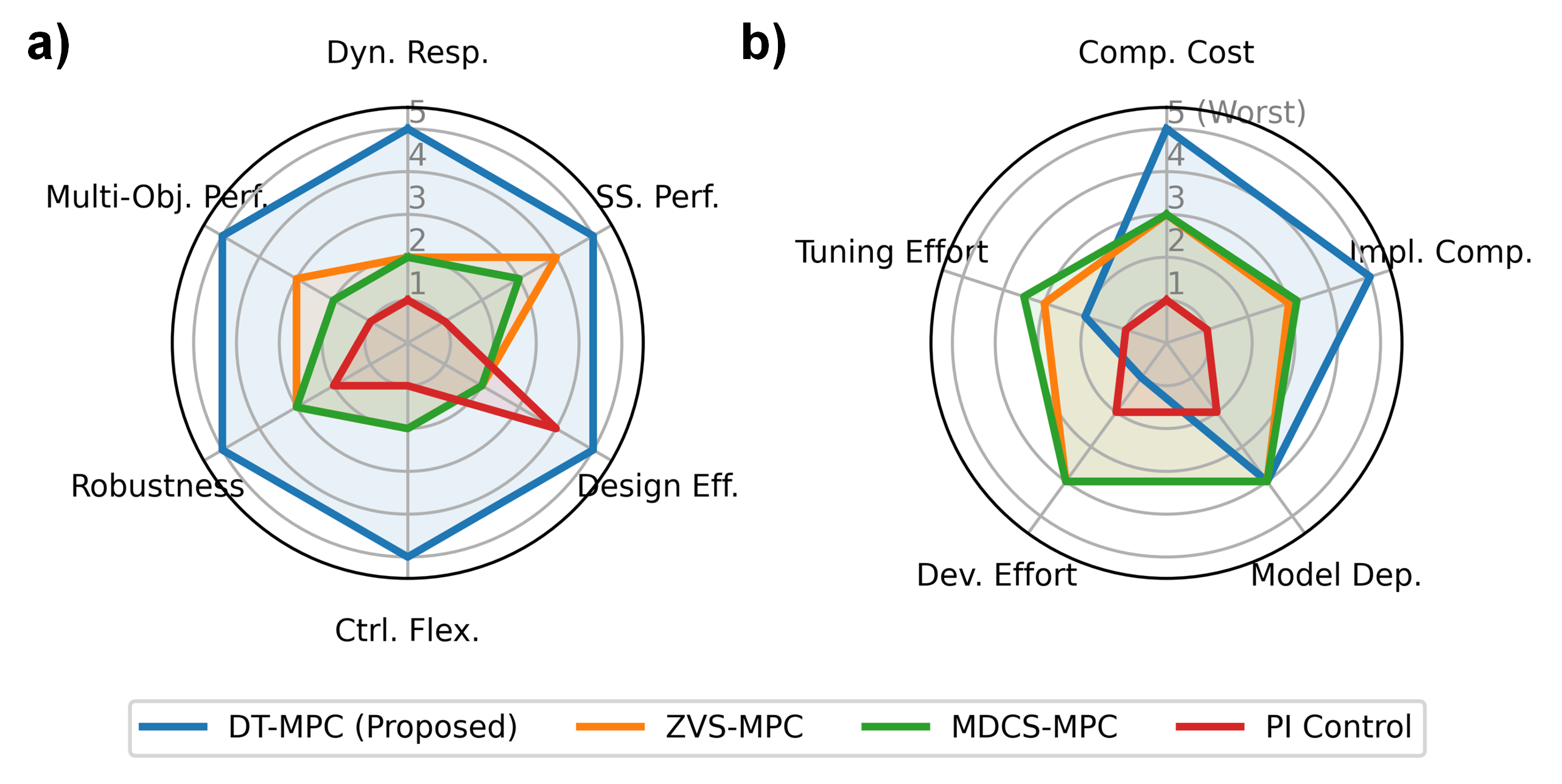} 
    \vspace{-0.8 cm}
    \caption{Multi-dimensional comparison of the proposed DT-MPC against benchmark controllers. (a) The performance and flexibility radar chart. (b) The cost and complexity radar chart illustrates the trade-offs. Abbreviations: Dyn. Resp. (Dynamic Response), SS. Perf. (Steady-State Performance), Design Eff. (Design Efficiency), Ctrl. Flex. (Control Flexibility), Multi-Obj. Perf. (Multi-Objective Performance), Comp. Cost (Computational Cost), Impl. Comp. (Implementation Complexity), Model Dep. (Model Dependency), Dev. Effort (Development Effort), and Tuning Effort.}
\label{fig:radar_summary}
\end{figure}

The experimental results have demonstrated the performance advantages of the DT-MPC framework. This is not an incremental advance but a consequence of a fundamentally new design paradigm. With the multi-dimensional comparison in Fig.~\ref{fig:radar_summary} as a guide, this section synthesizes the experimental findings to discuss the foundational advantages, inherent trade-offs, and broader implications of this approach.

The most profound contribution is a paradigm shift from manual derivation to automated synthesis. This is quantitatively demonstrated by the contrast in `Dev. Effort` and `Design Eff.` in Fig.~\ref{fig:radar_summary}. Traditional MPC design is fundamentally bottlenecked by the need for an expert to manually derive complex, often piecewise, analytical models. To provide a concrete example, several co-authors independently derived the analytical model for the DAB converter used in this study; this manual process, requiring deep domain expertise, averaged 46 man-hours. In stark contrast, the DT-MPC framework automates this entire process. The total time to generate a deployable controller from a completed netlist was approximately 117 minutes, which includes the human effort of writing the circuit netlist (approx. 40 minutes) followed by the automated steps of NSP training (37 minutes) and final deployment onto the FPGA (23 minutes). Compared to the traditional approach, this represents a reduction in engineering time of over 95\%, transforming advanced control design from a bespoke craft into a scalable, systematic process.

The second foundational advantage is a new frontier in control performance, unlocked through unconstrained online optimization. The experiments confirm that the exceptional `Dynamic Response` of DT-MPC (Fig.~\ref{fig:radar_summary}(a)) stems directly from its independence from analytical models. Conventional methods like ZVS-MPC are often built on steady-state assumptions that constrain their dynamic degrees of freedom and compromise performance during large disturbances. By performing real-time exploration on a high-fidelity model, DT-MPC synergistically utilizes all control variables. This allows it to achieve what was previously a difficult compromise: the superior dynamic response of a fully unconstrained controller combined with the steady-state optimality (`SS. Perf.`) of a specialized, hand-designed method.

These gains in efficiency and performance are enabled by an intentional trade-off: a higher initial implementation and computational complexity, as honestly depicted by the `Comp. Cost` and `Impl. Comp.` scores in Fig.~\ref{fig:radar_summary}(b). The framework's computational cost stems from the neural-surrogate predictor, which in turn necessitates a more sophisticated implementation based on a heterogeneous computing architecture that integrates AI-inference engines with real-time control logic.

Crucially, this perceived hurdle is largely mitigated by the modern hardware and software ecosystem. Commercial FPGA System-on-Chips (SoCs), for instance, provide a single-die solution combining ARM cores, programmable logic, and dedicated AI engines. The low-level integration of these components is abstracted away by mature, high-level synthesis (HLS) toolchains, which shifts the engineering challenge. Instead of requiring a deep, theoretical understanding of converter modeling, the task becomes a more manageable, one-time setup of an established deployment workflow. 

We contend that this is a highly favorable trade-off, as it eliminates the recurring, error-prone, and non-scalable bottleneck of manual derivation, effectively democratizing the design of high-performance controllers. Furthermore, this one-time implementation barrier is set to diminish, driven by parallel advancements in both hardware and algorithms. As Moore's Law continues to improve edge computing, future work will also focus on reducing computational demands at the algorithmic level by incorporating techniques such as model order reduction and causal inference.

In summary, the DT-MPC framework is not merely a higher-performing controller; it is a fundamental re-imagining of the control design process. It trades a one-time, automatable implementation effort for a drastic reduction in recurring engineering design time and a significant leap in control performance and flexibility.

\section{Conclusion} 

This paper presents and validates a derivation-free automated DT-MPC synthesis framework to address the critical dependence of conventional MPC on manual analytical modeling. The cornerstone of this framework is the programmatic generation of a high-fidelity digital twin model from a circuit netlist, combined with a neural surrogate predictor and a simplex search optimizer  to enable real-time control. Comprehensive performance evaluation and experimental validation on a DAB converter have demonstrated that the proposed NSP achieves faster-than-real-time prediction speed with high accuracy, the SSO shows superior convergence and optimization performance in high-dimensional continuous spaces over traditional methods, and the complete DT-MPC system exhibits far superior dynamic response compared to benchmark controllers while matching and even exceeding their steady-state performance. More significantly, this work represents a paradigm shift in control design from model derivation and manual customization to data-driven synthesis and automatic generation. By automating the burdensome modeling process, it dramatically reduces the development barrier for advanced control strategies, allows engineers to focus on high-level performance objective definition, and offers a viable and scalable path forward for tackling the control challenges of advanced power electronic systems.

\section*{Acknowledgment}

Yangbin Zeng  and Hong Li acknowledge the financial support of the NSFC under Grants 52577197 and 52237008. Leopoldo Franquelo and Sergio Vazquez acknowledge the financial support of Project PID2023-1522920B-100 funded by MCIN/AEI (10.13039/501100011033) and by ERDF/EU.

\section*{Appendix} \label{appendix:proof} 

In this section, we provide a proof for the global convergence of the proposed simplex-based search algorithm to a stationary point.

\begin{theorem}[Convergence to Stationary Points]
Let the cost function $J: \mathbb{R}^n \to \mathbb{R}$ be continuously differentiable ($C^1$). Assume the sequence of iterates $\{U_k\}$ generated by the algorithm remains within a bounded set $\mathcal{D}$, and that the gradient $\nabla J$ is Lipschitz continuous on $\mathcal{D}$, i.e., there exists a constant $L > 0$ such that for all $U, V \in \mathcal{D}$:
$$
\|\nabla J(U) - \nabla J(V)\| \le L \|U - V\|.
$$
Then, any cluster point of the sequence $\{U_k\}$ is a stationary point of $J$.

\end{theorem}

\begin{proof}
We proceed by contradiction. Assume that there exists a subsequence $\{U_{k_j}\}$ that converges to a point $U^*$, but $U^*$ is not a stationary point, i.e., $\nabla J(U^*) \neq 0$.

Since $\nabla J$ is continuous, for all sufficiently large $j$, we must also have $\nabla J(U_{k_j}) \neq 0$. According to the algorithm's procedure, when the standard simplex transformations fail to produce a point satisfying the sufficient decrease condition, a search is performed along the directions of a positive basis $\mathcal{V}^+_{k_j} = \{v_1, \dots, v_{n+1}\}$ centered at the best vertex $U_{k_j}$. By the property of the positive basis, there must exist at least one vector $v_{k_j} \in \mathcal{V}^+_{k_j}$ and a constant $\delta_0 > 0$ (independent of $j$) such that this vector provides a descent direction:
\begin{equation}
\nabla J(U_{k_j})^{\!\top} v_{k_j} \le -\delta_0.
\label{eq:descent_condition}
\end{equation}
By the Descent Lemma (a consequence of Taylor's theorem for functions with a Lipschitz continuous gradient), for a step of size $h_{k_j}$ along this direction, we have:
\begin{equation}
J(U_{k_j} + h_{k_j}v_{k_j}) \le J(U_{k_j}) + h_{k_j} \nabla J(U_{k_j})^{\!\top} v_{k_j} + \frac{1}{2}L \|h_{k_j}v_{k_j}\|^2.
\end{equation}
Substituting ~\eqref{eq:descent_condition} into the inequality above gives:
\begin{equation}
J(U_{k_j} + h_{k_j}v_{k_j}) \le J(U_{k_j}) - h_{k_j}\delta_0 + \frac{1}{2}Lh_{k_j}^2 \|v_{k_j}\|^2.
\end{equation}
This can be rearranged to show the guaranteed decrease in the cost function:
\begin{equation}
J(U_{k_j}) - J(U_{k_j} + h_{k_j}v_{k_j}) \ge h_{k_j}\delta_0 - \frac{1}{2}Lh_{k_j}^2 \|v_{k_j}\|^2.
\label{eq:guaranteed_decrease}
\end{equation}
The algorithm's acceptance criterion for a new point, however, is the sufficient decrease condition:
\begin{equation}
J(U_{\text{new}}) \le J(U_{k_j}) - \sigma h_{k_j}^2,
\label{eq:sufficient_decrease}
\end{equation}
where $\sigma > 0$ is a predefined constant. A successful step is thus found if the guaranteed decrease in ~\eqref{eq:guaranteed_decrease} is at least as large as the required decrease in ~\eqref{eq:sufficient_decrease}. That is, if:
$$
h_{k_j}\delta_0 - \frac{1}{2}Lh_{k_j}^2 \|v_{k_j}\|^2 \ge \sigma h_{k_j}^2.
$$
Dividing by $h_{k_j}$ (which is positive) and rearranging the terms, we find that a successful step is guaranteed to exist as long as the step size $h_{k_j}$ is sufficiently small:
\begin{equation}
h_{k_j} \le \frac{\delta_0}{\sigma + \frac{1}{2}L\|v_{k_j}\|^2}.
\label{eq:h_condition}
\end{equation}
Here lies the contradiction. The premise for the subsequence $\{U_{k_j}\}$ is that these are the iterations where the algorithm failed to find an improving point via transformations and had to resort to the positive basis search. The algorithm only reduces the search scale $h_k$ to $h_{k+1} = \rho h_k$ when all probes along the positive basis directions fail to satisfy the sufficient decrease condition. However, ~\eqref{eq:h_condition} shows that as the algorithm repeatedly reduces $h$, the step size $h_{k_j}$ will inevitably become small enough to satisfy the condition. At that point, a successful step would have been taken, and $U_{k_j}$ would not have been declared quasiminimal, nor would $h$ have been further reduced. This contradicts the assumption that we can form a subsequence where $h_{k_j}$ can be made arbitrarily small while failing to find a descent direction.

Therefore, the initial assumption must be false, and we must have $\nabla J(U^*) = 0$. This completes the proof.

\end{proof}

%

\bibliographystyle{IEEEtran}
\bibliography{Bibliography/BIB_xx-TPEL-xxxx}


\vfill

\end{document}